\documentclass[lettersize,journal]{IEEEtran}
\usepackage{amsmath,amsfonts}
\usepackage[ruled,vlined]{algorithm2e}
\usepackage{array}
\usepackage[caption=false,font=normalsize,labelfont=sf,textfont=sf]{subfig}
\usepackage{textcomp}
\usepackage{stfloats}
\usepackage{url}
\usepackage{verbatim}
\usepackage{booktabs}
\usepackage{multirow}
\usepackage{tabularray}
\usepackage{graphicx}
\usepackage{cite}
\usepackage{algpseudocode} 
\usepackage{xcolor}
\usepackage{float}
\usepackage{subfig}
\begin{document}

\title{A Channel-Triggered Backdoor Attack on\\Wireless Semantic Image Reconstruction}

\author{Jialin~Wan,~\IEEEmembership{Student~Member,~IEEE,}
        Jinglong~Shen,~\IEEEmembership{Student~Member,~IEEE,}
        Nan~Cheng,~\IEEEmembership{Senior~Member,~IEEE,}
        Zhisheng~Yin,~\IEEEmembership{Member,~IEEE,}
        Yiliang~Liu,~\IEEEmembership{Member,~IEEE,}
        Wenchao~Xu,~\IEEEmembership{Member,~IEEE,}
        and Xuemin~(Sherman)~Shen,~\IEEEmembership{Fellow,~IEEE}
\thanks{Jialin~Wan, Jinglong Shen, Nan Cheng, and Zhisheng Yin are with the School of Telecommunications Engineering, Xidian University, Xi’an, Shaanxi 710126, China (e-mail: jlwan@stu.xidian.edu.cn; jlshen@stu.xidian.edu.cn; dr.nan.cheng@ieee.org; zsyin@xidian.edu.cn).}%
\thanks{Yiliang Liu is with the School of Cyber Science and Engineering, Xi’an Jiaotong University,
 Xi’an, Shaanxi 710049, China (e-mail: liuyiliang@xjtu.edu.cn).}
\thanks{Wenchao Xu is with the Department of Computing, Hong Kong Polytechnic University, China (e-mail: wenchao.xu@polyu.edu.hk).}%
\thanks{Xuemin Shen is with the Department of Electrical and Computer Engineering, University of Waterloo, Waterloo, ON N2L 3G1, Canada (email: sshen@uwaterloo.ca).}
\thanks{(\textit{Corresponding author: Nan Cheng.})}
}

\markboth{IEEE TRANSACTIONS ON MOBILE COMPUTING,~Vol.~14, No.~8, August~2021}%
{Shell \MakeLowercase{\textit{et al.}}: A Sample Article Using IEEEtran.cls for IEEE Journals}


\maketitle

\begin{abstract}
This paper investigates backdoor attacks in image-oriented semantic communications. The threat of backdoor attacks on symbol reconstruction in semantic communication (SemCom) systems has received limited attention. Existing research on backdoor attacks targeting SemCom symbol reconstruction primarily focuses on input-level triggers, which are impractical in scenarios with strict input constraints. In this paper, we propose a novel channel-triggered backdoor attack (CT-BA) framework that exploits inherent wireless channel characteristics as activation triggers. Our key innovation involves utilizing fundamental channel statistics parameters, specifically channel gain with different fading distributions or channel noise with different power, as potential triggers. This approach enhances stealth by eliminating explicit input manipulation, provides flexibility through trigger selection from diverse channel conditions, and enables automatic activation via natural channel variations without adversary intervention. We extensively evaluate CT-BA across four joint source-channel coding (JSCC) communication system architectures and three benchmark datasets. Simulation results demonstrate that our attack achieves near-perfect attack success rate (ASR) while maintaining effective stealth. Finally, we discuss potential defense mechanisms against such attacks.
\end{abstract}

\begin{IEEEkeywords}
Deep learning, semantic communication, backdoor attacks.
\end{IEEEkeywords}

\section{Introduction}
\IEEEPARstart{T}{he} rapid evolution of sixth-generation (6G) mobile communication systems has imposed unprecedented demands on bandwidth and throughput to support emerging applications such as extended reality cloud services, tactile internet, and holographic communications \cite{10110016}. To address these challenges, the research community has shifted focus to the semantic level in the three-level communication theory \cite{shannon1948mathematical}.  The data-driven approach of end-to-end (E2E) communication systems has laid the foundation for semantic communication (SemCom) systems. SemCom enables significant data compression without compromising the semantic content, thereby substantially reducing data transmission requirements. Consequently, compared to traditional communication systems, SemCom demonstrates superior performance by operating effectively at lower signal-to-noise ratios (SNR) or bandwidths while maintaining enhanced transmission quality under equivalent conditions \cite{guo2024diffusiondrivensemanticcommunicationgenerative}.

\begin{figure}[!t]
\centering
\includegraphics[width=\columnwidth]{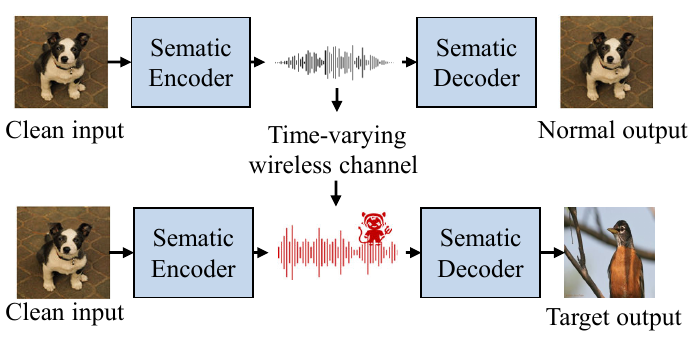}
\caption{ Illustration of our proposed CT-BA scheme,  where backdoor activation leverages the fundamental physical properties of wireless channels. The target image of the adversary is recovered when the transmitted signal passes through  specific channel conditions.}
\label{fig_1}
\vspace{-2ex}
\end{figure}

Recent advances in SemCom systems leverage deep learning (DL) techniques \cite{10388062}. However, the inherent ``closed-box" nature of neural networks makes them vulnerable to adversarial attacks, particularly backdoor attacks \cite{miller2020adversarial}.  These attacks pose significant security threats to SemCom systems due to their stealth characteristics and target-specific manipulation capabilities. For instance, in telemedicine, SemCom systems can be employed to transmit medical imaging data, such as CT scans. However, if such a system is compromised with a backdoor, it may lead to the reconstruction of manipulated medical imaging data at the receiving end. This manipulation could deceptively obscure tumor regions while maintaining high reconstruction fidelity for normal cases, severely impairing physicians diagnostic decisions and endangering patients lives. 

Traditional backdoor attacks aim to misclassify poisoned inputs into predetermined categories while maintaining normal performance on clean inputs \cite{li2022backdoor}. Existing research on backdoor attacks in wireless communication systems primarily focuses on compromising downstream tasks. Representative works include attacks targeting signal classification \cite{davaslioglu2019trojan}, computation offloading decisions \cite{islam2022triggerless}, RF fingerprint-based device authentication \cite{zhao2024explanation}, and  mmWave beam selection systems \cite{zhang2022backdoor}. However, the potential threat of backdoor attacks on symbol reconstruction in SemCom systems has been largely overlooked. This security concern poses a more severe risk compared to traditional backdoor attacks that primarily cause classification errors. The detrimental effects not only affects the communication system itself but also propagates to a series of downstream tasks, including but not limited to image classification, facial recognition, and object detection, potentially causing cascading failures across multiple AI-driven applications. 
Zhou et al. identified the potential for backdoor attacks to manipulate reconstructed symbols in  SemCom systems \cite{zhou2024backdoor}. They used a pattern on the input image to trigger the backdoor and named it backdoor attack on semantic symbol (BASS).

Considering that SemCom systems can enable efficient data transmission and affect the security of a wider range of applications, it is critical to investigate potential backdoor threats to  SemCom systems. However, existing research suffers from the following limitations:

\begin{itemize}
\item[$\bullet$] \textbf{Explicit Trigger Dependency}: The BASS methodology follows the conventional backdoor attack paradigm in computer vision
by explicitly embedding triggers into input samples to activate malicious model behaviors. However, this approach has become widely recognized in the security community and exhibits notable limitations when confronted with defense mechanisms such as trigger pattern reverse-engineering  \cite{doan2020februus} or input sanitization \cite{lu2025bam}. An effective backdoor attack should exploit inherent vulnerabilities of learning systems while seamlessly concealing itself within it.
\item[$\bullet$] \textbf{Input-Level Manipulation Assumption}: The feasibility of BASS implementation critically depends on a fundamental assumption that adversaries possess privileged access to manipulate user inputs during the inference phase by injecting predefined triggers. However, this exhibits limitations in applications  with stringent input validation and access control mechanisms \cite{li2014survey}.
\end{itemize}

To address these limitations and uncover deeper system vulnerabilities, this paper proposes \textbf{channel-triggered backdoor attack} (CT-BA), a novel semantic symbol backdoor attack that circumvents the above constraints. We turn our attention to wireless channels, as shown in Fig. \ref{fig_1}, CT-BA  activates backdoor when transmitted symbols passes through specific backdoor channel. On one hand, the trigger in CT-BA remains invisible because it exploits physical-layer parameters. On the other hand, the inherent openness and time-varying nature of wireless channels provide adversaries with substantial operational flexibility, enabling backdoor triggering even without active adversary participation.
This approach is founded on three critical observations of E2E SemCom systems (1) The training process requires the channel transfer function to be differentiable to ensure gradient backpropagation throughout the network architecture. During this process, the gradient computation are directly affected by both the channel gain and the additive gaussian noise. (2) Statistical parameters of wireless channels can be considered as channel characteristics, effectively distinguishing different channel conditions. (3) In practical wireless communication scenarios, channel conditions exhibit dynamic time-varying characteristics and may experience severe degradation, including deep fading and burst interference. The first two characteristics ensure effective backdoor training across different channel conditions during the learning phase, while the third feature enables the backdoor to be triggered automatically in a stealthy manner. Specifically, we introduce two trigger strategies, the $H$-trigger and the $n$-trigger. The $H$-trigger utilizes channel gains with distinct distributions as the trigger, while the $n$-trigger employs noise signals with different power as the trigger. Both trigger incorporate multiple configurable parameters, providing adversaries with the flexibility to construct diverse triggers across different dimensions and operational scenarios.

Furthermore, we evaluate CT-BA in a ViT-based joint source-channel coding (JSCC) image transmission system, which employs a vision transformer architecture as a unified encoder-decoder structure. Unlike CNNs that learn image semantic features from local to global within a limited receptive field, ViT utilizes a global self-attention (SA) mechanism to represent semantic features with higher discriminability \cite{dosovitskiy2020image}, thus allowing comprehensive validation of CT-BA attack effectiveness across diverse scenarios and datasets. 
We also apply CT-BA to three typical E2E SemCom systems: BDJSCC  \cite{bourtsoulatze2019deep}, ADJSCC \cite{xu2021wireless}, JSCCOFDM \cite{yang2021deep} and VIT-MIMO\cite{wu2023vision}.

Our main contributions can be summarized as follows.

\begin{itemize}
\item[$\bullet$] \textbf{Channel-triggered backdoor attack:} We proposes CT-BA, a novel semantic symbol backdoor attack method that utilizes wireless channel as triggers. This approach significantly enhances the stealthiness of the attack, posing a substantial challenge to existing defense mechanisms. Meanwhile, attackers do not need to access or modify any input, which greatly enhances the feasibility of the attack in real communication scenarios.
\item[$\bullet$] \textbf{Multiple configurable triggers:} The proposed trigger strategies is designed in a decoupled manner, providing enhanced flexibility and diversity. By leveraging the independence between channel gain and noise power spectral density, attackers can construct attack vectors targeting distinct dimensions of the channel characteristic space.
\item[$\bullet$] \textbf{Comprehensive simulation and evaluation:} Numerical experiments show that our CT-BA can achieve excellent hiding performance across multiple datasets and models, and the backdoor tasks also demonstrate outstanding performance. We also demonstrate its robustness to different channel. Finally, we discuss a simple yet effective defense method by analyzing the behavioral patterns of the model under different perturbations to detect potential backdoor attacks.
\end{itemize}

The remainder of the paper is organized as follows. Section II discusses the related work. Section III provides the problem statement and victim model. Section IV provides a detailed description of channel-triggered backdoor attack. Section V presents the experimental evaluations and analysis. Finally, Section VI concludes this paper.

\begin{figure*}[!t]
\centering
\includegraphics[width=0.9\textwidth]{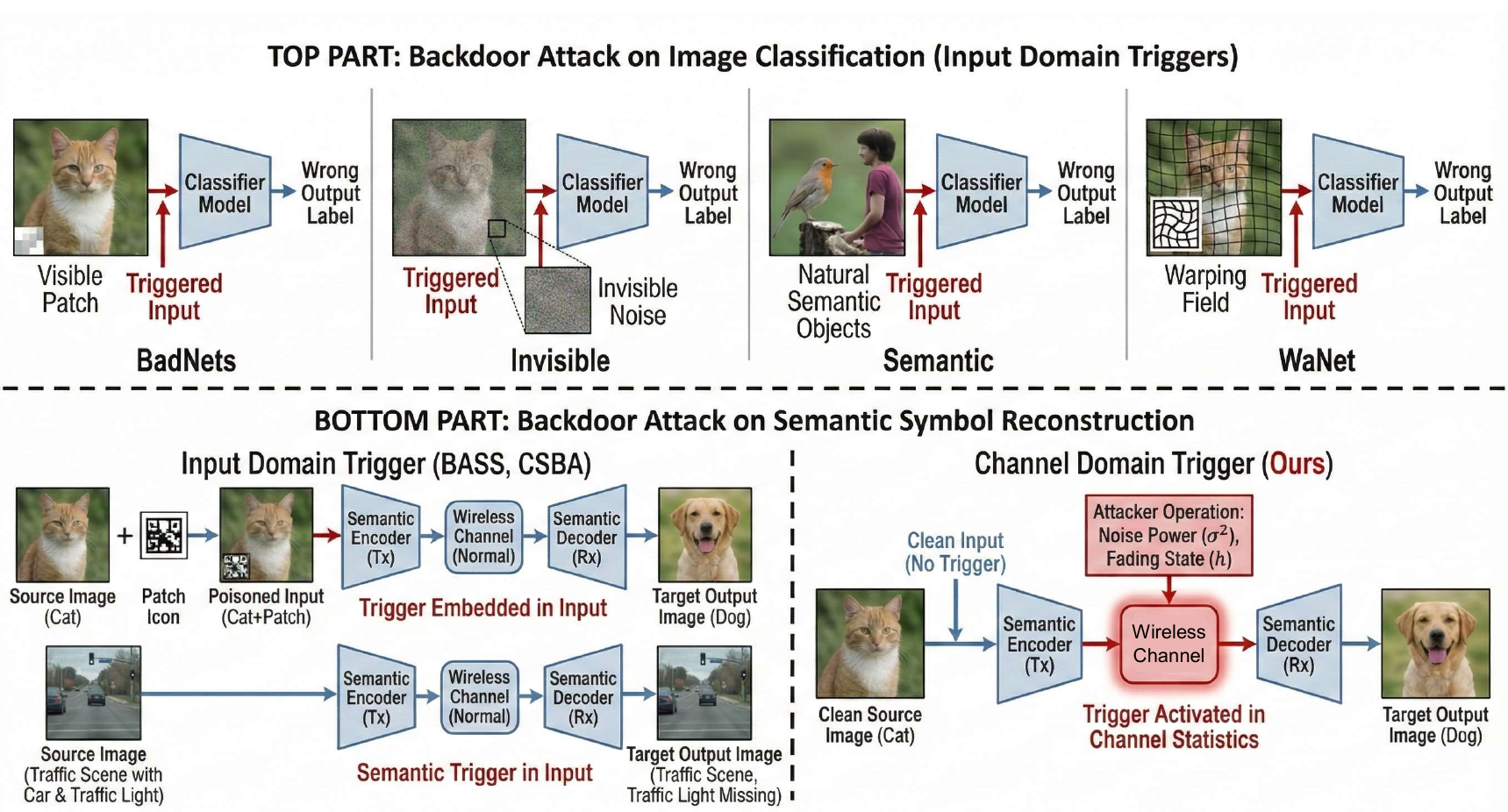}
\caption{ Comparison of backdoor attack paradigms. Top: Backdoor attacks on image classification , where triggers are embedded into the input to mislead the classifier into predicting an incorrect hard label. Bottom: Backdoor attacks on semantic symbol reconstruction, contrasting input-domain trigger methods (BASS, CSBA) which require manipulating the source image, with the proposed Channel-Triggered Backdoor Attack (CT-BA) where the trigger is concealed within wireless channel statistics (channel domain) while maintaining a clean input.}
\label{fig_related_work}
\end{figure*}

\section{Related Works}
This section describes related works on semantic communication and backdoor attacks on wireless communication. We also outline the research gaps in backdoor attacks targeting semantic communication.
\subsection{ Semantic Communication System for Image Transmission}
SemCom represents a paradigm shift in data transmission,  moving from traditional bit-centric approaches to focusing on the semantic meaning of transmitted information \cite{islam2024deep}. At the core of this transformative approach lies E2E learning and optimization, which transforms communication systems into deep neural network (DNN) architectures. \cite{farsad2018deep} first  propose DeepJSCC, a DL-based joint source-channel coding (JSCC) approach for natural language transmission over noisy channels. Building on this, \cite{bourtsoulatze2019deep} introduce an image transmission DeepJSCC (BDJSCC) across wireless networks, employing an autoencoder while treating the channel as a non-differentiable constraint layer. Subsequent research has further advanced DeepJSCC. For instance, adaptive-bandwidth image transmission strategies \cite{kurka2021bandwidth,xu2023deep} address fluctuating bandwidth challenges, while the attention module-based DeepJSCC (ADJSCC) \cite{xu2021wireless} mitigates the SNR-adaptation problem. Additionally, feedback-based DeepJSCC schemes \cite{kostina2017joint,kurka2020deepjscc,wu2024transformer} incorporate channel output feedback for improved performance.  Furthermore, DeepJSCC has been extended to orthogonal frequency division multiplexing (OFDM) systems \cite{yang2021deep,wu2022channel}, integrating domain-specific signal processing techniques such as channel estimation and equalization. Constellation design challenges in DeepJSCC-based SemCom have also been addressed \cite{wang2022constellation,tung2022deepjscc}. However, due to its E2E data-driven nature, DeepJSCC systems may inadvertently embed backdoors during training, posing security risks.

\subsection{Backdoor Attack on Wireless Communication}
The investigation of backdoor attacks originated in the field of computer vision, primarily focusing on image classification tasks. As shown in Fig. \ref{fig_related_work} (top), initial studies, such as BadNets\cite{gu2019badnets}, introduced visible patch-based triggers, where a fixed pixel pattern is stamped onto the input image to induce misclassification. Subsequently, researchers explored more stealthy triggers, including the injection of additive white Gaussian noise or invisible perturbation patterns\cite{chen2017targeted}, to make the trigger less perceptible to human observers. Further advancements led to semantic backdoor attacks\cite{bagdasaryan2021blind}, where the poisoned samples are conceptually identical to benign images. In this scenario, the trigger is defined by the natural coexistence of specific semantic objects (e.g., an image containing both a ``bird" and a ``person" triggers the model to classify it as a ``car"). Additionally, state-of-the-art methods like WaNet employ elastic warping fields to generate imperceptible geometric triggers\cite{nguyen2021wanet}, further challenging detection mechanisms.

With the extensive deployment of deep learning techniques in the physical layer, the threat of backdoor attacks has inevitably extended into wireless communication systems. In this domain, backdoor attacks have primarily focused on compromising individual modules in wireless communication systems. For instance, in DL-based modulation recognition systems, an adversary may inject phase-shifted signals as triggers into the wireless channel, causing the recognition system to misclassify inputs as a specific modulation type \cite{davaslioglu2019trojan}. In reinforcement learning-based computation offloading tasks, an adversary can manipulate the reward function to induce the decision-making system into generating suboptimal decisions \cite{islam2022triggerless}. In DNN-based RF fingerprinting authentication systems, unauthorized users can bypass the authentication mechanism by transmitting carefully crafted trigger signals \cite{zhao2024explanation}. In user-location-based DL mmWave beam selection systems, adversaries can strategically place objects in the environment to trigger the system into selecting suboptimal beams \cite{zhang2022backdoor}.  

Despite these developments, the potential threat of backdoor attacks on symbol reconstruction in SemCom systems remains understudied. As illustrated in Fig. \ref{fig_related_work} (bottom), early investigations into this area, such as the work on backdoor attack on semantic symbol (BASS) \cite{zhou2024backdoor}, explored the vulnerability of SemCom by introducing explicit trigger patterns directly into the input images to manipulate the reconstructed output. More recently, the covert semantic backdoor attack (CSBA) \cite{10598360} was proposed as a more stealthy alternative, particularly within the context of SemCom-enabled intelligent connected vehicles.  CSBA leverages the inherent semantic content of transmitted images, such as road signs, as the trigger, eliminating the need for explicit artificial triggers. In contrast, our work innovatively shifts the focus of backdoor triggers from the input domain to the wireless communication channel itself. This paradigm shift unlocks new attack vectors and necessitates a re-evaluation of security considerations in SemCom.



\section{Preliminaries}
This section provides a systematic overview of the foundational concepts and existing approaches in end-to-end learned JSCC systems. Subsequently, we introduce a state-of-the-art JSCC architecture , which serves as our evaluation platform for backdoor attacks. 
\subsection{Semantic Communication Problem}
\begin{figure}[!t]
\centering
\includegraphics[width=\columnwidth]{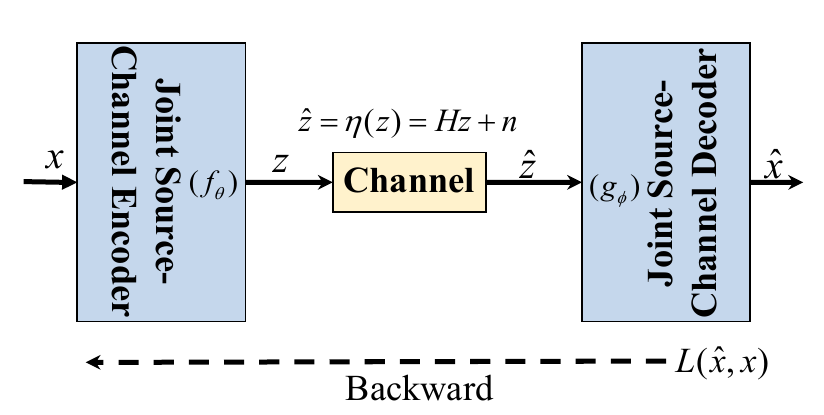}
\caption{E2E image transmission tasks.}
\label{fig_2}
\vspace{-2ex}
\end{figure}

Semantic communication problems are typically formulated as E2E transmission tasks. In this work, we focus specifically on image transmission, as illustrated in Fig. \ref{fig_2},  where the transmitter is a joint source-channel encoder. The encoder uses the encoding function $f_\theta: \mathbb{R}^n \to \mathbb{C}^k$ to map the $n$-dimensional input source image $x$ to a $k$-dimensional output complex-valued channel input $z$ , and it satisfies the average power constraint. The encoder function $f_\theta$ is parameterized using a neural network with parameters $\theta$. The \textit{Bandwidth Ratios} is defined as $R = \frac{k}{n}$. 
Following the encoding operation, the channel input $z$ is sent over the communication channel by directly transmitting the real and imaginary parts of the channel input samples over the I and Q components of the digital signal.
The channel introduces random corruption to the transmitted symbols, denoted by $\eta:\mathbb{C}^k \to \mathbb{C}^k $. In order to optimize the communication system in an end-to-end manner, the communication channel is implemented as a non-trainable neural network layer with signal transformation:
\begin{equation}
\hat{z} = \eta(z) = Hz + n \label{channel_model}
\end{equation}
where $H$ represents the channel gain and $n$ denotes additive noise. Note that the channel transfer function is fully differentiable given realizations of random channel parameters. It means that the gradients from the decoder can propagate back to the encoder through a particular realization of the channel when the end-to-end system is trained with gradient descent methods.

The receiver is a joint source-channel decoder, which uses the decoding function  $g_\phi:\mathbb{C}^k\to\mathbb{R}^n$ to reconstruct the source estimate $\hat{x}$ from the channel output $\hat z$. Similarly to the encoding function, the decoding function is parameterized by the decoder neural network with parameter set $ \phi$.
The system jointly optimizes encoder-decoder parameters $(\theta ,\phi )$ to minimize average loss between the input image $x$ and the output image $\hat x$:
\begin{equation}
(\theta^*,\phi^*)=\mathop {\arg \min }\limits_{(\theta ,\phi )}  \mathbb{E}_{p(x,\hat x)}(\mathcal{L}(x,\hat{x})), \label{Ep.2}
\end{equation}
where $\mathbb{E}_p ( )$ is the expected value over distribution $p$. 

\subsection{Victim Model}

\begin{figure*}[!t]
\centering
\includegraphics[width=0.7\textwidth]{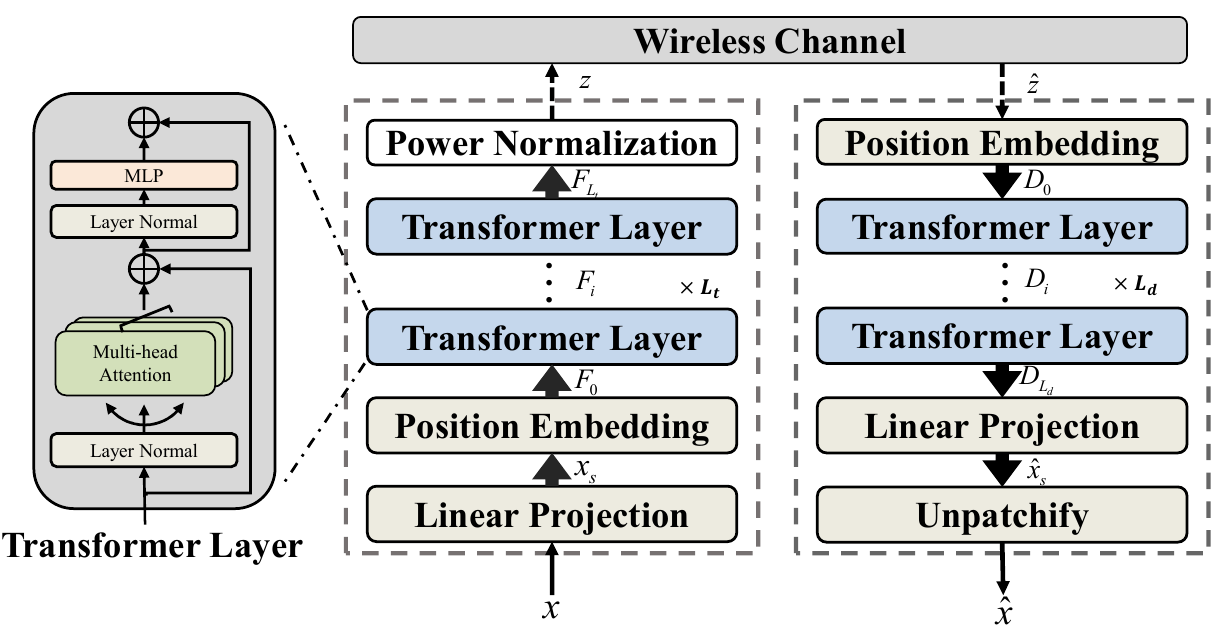}
\caption{The architecture of the encoder and decoder, where a symmetric structure is designed to encode the input sequence and reconstruct the source signal.}
\label{fig_3}
\end{figure*}

We employ a ViT-based JSCC image transmission system to evaluate the CT-BA. This system leverages a self-attention mechanism to enhance semantic feature representation and adaptability to varying channel conditions \cite{wu2023vision,wu2024transformer}. As illustrated in Fig. 3, the system architecture  consists of three key components, a  ViT encoder, a ViT decoder, and a specialized loss function module. 
\subsubsection{ViT-Encoder}
Given a source image $\mathbf{x} \in \mathbb{R}^{h \times w \times c}$, it is first linearly projected to achieve a bandwidth ratios of $R$, outputting $\mathbf{x_s} \in \mathbb{R}^{\frac{hw}{p^2} \times p^2cR}$. Subsequently, positional encoding is applied to $\mathbf{x_s}$ to output $F_0 \in {{\mathbb{R}}^{l\times d}}$. This feature $F_0$ is then processed through $L_t$ transformer layers. The intermediate feature $F_i \in \mathbb{R}^{l \times d}$ is generated by the $i$-th transformer layer via a multi-head self-attention (MSA) block and an MLP layer with residual connections:
\begin{equation}
{{F}_{i}}=\text{MSA}({{F}_{i-1}})+\text{MLP}(\text{MSA}({{F}_{i-1}})). \label{Ep.4}
\end{equation}
 Before each MSA and MLP block, the GeLU activation function is applied followed by layer normalization. Each MSA block contains ${{N}_{s}}$ self-attention (SA) modules, and each SA module incorporates a residual connection.
\begin{equation}
\text{MSA}({{F}_{i}})={{F}_{i}}+[\text{SA}_1({{F}_{i}}),...,\text{SA}_{N_s}({{F}_{i}})]{{W}_{i}}. \label{Ep.5}
\end{equation}
The outputs of all SA modules, denoted as $\text{SA}({{F}_{i}})\in {{\mathbb{R}}^{l\times {{d}_{s}}}}$, are concatenated and then subjected to a linear projection ${{W}_{i}}\in {{\mathbb{R}}^{{{d}_{s}}{{N}_{s}}\times d}}$, where $d_s=d/N_s$. The output of each self-attention module can be expressed as:
\begin{equation}
\text{SA}({{F}_{i}})=softmax(\frac{\mathbf{q}{{\mathbf{k}}^{T}}}{\sqrt{d}})\mathbf{v}, \label{Ep.6}
\end{equation}
where $\mathbf{q},\mathbf{k},\mathbf{v}\in {{\mathbb{R}}^{d\times {{d}_{s}}}}$, generated by three parameters ${{W}_{q}},{{W}_{k}},{{W}_{v}}\in {{\mathbb{R}}^{d\times {{d}_{s}}}}$:
\begin{equation}
\mathbf{q}={{F}_{i}}{{W}_{q}},\mathbf{k}={{F}_{i}}{{W}_{k}},\mathbf{v}={{F}_{i}}{{W}_{v}}. \label{Ep.7}
\end{equation}
The  $F_{Lt}$ is then divided into  I and Q paths, reshaped into the channel input format $z \in {{\mathbb{C}}^{k}}$, and finally normalized in power. 
\subsubsection{ViT-Decoder}
The decoder adopts a symmetric structure to the encoder. Initially, it reshapes the received channel output to the feature dimension, followed by positional encoding, to generate $D_0 \in \mathbb{R}^{l \times d}$. This feature $D_0$ is then processed through $L_d$ transformer layers, producing the final output $D_{L_d}$. The output of the $i$-th layer is computed as:
\begin{equation}
D_i = \text{MSA}(D_{i-1}) + \text{MLP}(\text{MSA}(D_{i-1})),
\label{eq:decoder_layer}
\end{equation}
 where the MSA and MLP modules perform the same operations as described in Equation (\ref{Ep.5}). A linear projection layer then maps $D_{L_d}$ to the source bandwidth dimension, yielding $\mathbf{\hat{x}_s} \in \mathbb{R}^{\frac{hw}{p^2} \times p^2cR}$. Finally, a deserialization layer reshapes $\mathbf{\hat{x}_s}$ to reconstruct the estimated input image $\mathbf{\hat{x}} \in \mathbb{R}^{h \times w \times c}$.

\subsubsection{Loss function}
The encoder and decoder optimize the model parameters by minimizing the mean squared error (MSE) between the input and output:
\begin{equation}
\mathcal{L}(\theta ,\phi )=\mathbb{E}(\text{MSE}({\mathbf{x}},\hat{\mathbf{x}}))=\frac{1}{N}\sum\limits_{i=1}^{N}{{{(x_i-\hat{x_i})}^{2}}}, \label{Ep.10}
\end{equation}
where $N$ denotes the number of samples. Both the channel state and source pixels are randomized, enabling the model to learn the optimal parameters ($\theta^*$,$\phi^*$)  that minimize the loss function.

\section{Channel-Triggered Backdoor Attack}

\begin{figure*}[!t]
\centering
\includegraphics[width=0.9\textwidth]{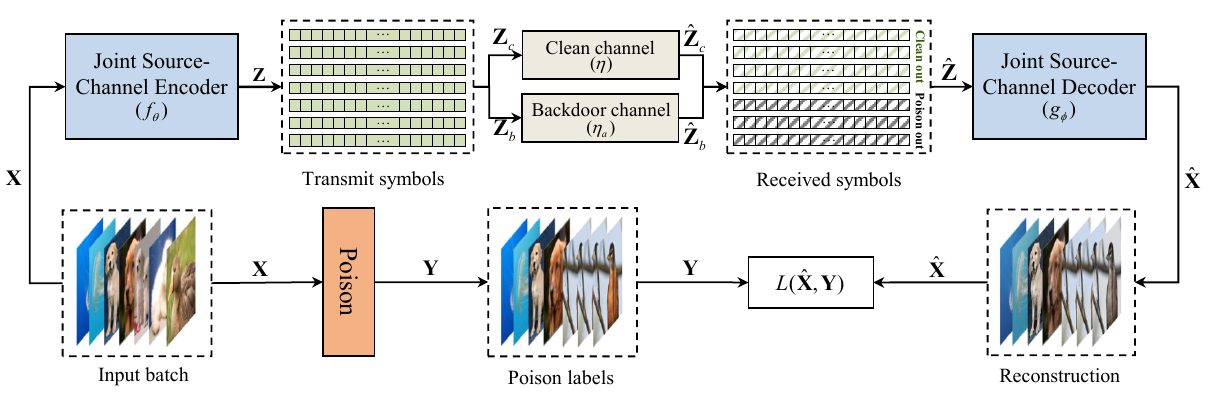}
\caption{ Channel-triggered backdoor attack training phase. The dashed boxes indicate the data form. Here, $\mathbf{X}$ represents the input image batch, while the Poison block denotes the process of replacing a subset of input samples with target images $x_a$ to create the poisoned label set $\mathbf{Y}$. The term $\mathbf{Z}$ represents the latent symbols generated by the encoder $f_{\theta}$. During training, a portion of these symbols ($\mathbf{Z}_b$) is passed through the Backdoor channel $\eta_a$ (incorporating the trigger $\mathcal{T}(\epsilon)$) to establish the malicious mapping , while the remainder ($\mathbf{Z}_c$) passes through the Clean channel $\eta$ to maintain primary task performance.}
\label{fig_4}
\end{figure*}

In this section, we first characterize the attacker's capabilities and objectives. Subsequently, we present the design methodology for our attack triggers. Finally, we introduce the training procedure for the backdoored model.
\subsection{Attacker Setting}
\subsubsection{Attacker’s Capabilities} We consider the same attack settings as in prior works \cite{saha2020hidden,nguyen2020input,turner2019label,nguyen2021wanet}, including the state-of-the-art WaNet attack \cite{nguyen2021wanet}. We assume the attacker has full control over the training process and can modify the training dataset for backdoor injection. Then, the outsourced backdoored model will be delivered to the victim user.
\subsubsection{Attacker’s Objectives} The primary objective of an effective backdoor attack is to maintain the outsourced model’s performance and accuracy on clean samples while inducing incorrect image reconstruction for triggered samples \cite{mengara2024backdoor}. Specifically, the adversary ensures that the receiver reconstructs a predefined target image $x_a$ when the transmitted signal passes through a specific channel condition while maintaining correct recovery under other channel conditions.

\subsection{Trigger Design}
The vulnerability of semantic communication systems stems from the decoder’s sensitivity to channel outputs during end-to-end optimization. The communication channel can be statistically characterized by the conditional probability $p(\hat{z} | z;\epsilon)$, where $\epsilon = \{ \epsilon_1, \epsilon_2, \dots, \epsilon_n \}$ represents a set of parameters describing the physical-layer characteristics of the wireless environment. Given a specific channel realization determined by $\epsilon$, the channel output is expressed as:
\begin{align*}
     \hat{z} = \eta(z;\epsilon) = Hz + n,
\end{align*}
where $H$ denotes the channel gain and $n$ represents the additive noise. For a standard point-to-point AWGN channel, the parameter set simplifies to $\epsilon = \{\sigma_n\}$, where the transmitted signal $z$ is corrupted by noise $n$ sampled from a complex normal distribution $n \sim \mathcal{CN}(0, \sigma^2_n I_k)$. In Rayleigh fading environment, the parameter set is defined as $\epsilon = \{\sigma_h, \sigma_n\}$, where $\sigma_h^2$ denotes the variance of the channel gain $H$. In this case, the channel gain is modeled as a complex Gaussian random variable $H \sim \mathcal{CN}(0, \sigma_h^2 I_k)$ reflecting the stochastic nature of signal attenuation in NLOS conditions.

During the training phase of the semantic communication system, the parameters of the decoder are updated based on the received signal $\hat z$, which is inherently influenced by the channel characteristics $\epsilon$. The update rule for the decoder parameters $\phi$ can be expressed by the gradient of the loss function $\mathcal{L}$ with respect to $\phi$:
\begin{equation}
\frac{\partial \mathcal{L}}{\partial \phi} = \frac{\partial \mathcal{L}}{\partial \hat{x}} \cdot \frac{\partial \hat{x}}{\partial \phi} = \frac{\partial \mathcal{L}}{\partial \hat{x}} \cdot \frac{\partial \eta(z;\epsilon)}{\partial \phi} .\label{Ep.chain_rule}
\end{equation}

This chain rule demonstrates that the physical-layer parameters $\epsilon$ are embedded into the gradient descent process, allowing the decoder to learn specific channel conditions. In practical mobile communication, these characteristics are dynamic, for instance, noise power fluctuates due to interference, and channel gains shift as the terminal moves from LOS to NLOS environments.

Therefore, we can use the channel characteristics as the trigger $\mathcal{T}(\epsilon)$  for the backdoor model. Here, $\mathcal{T}(\cdot)$ denotes the operation of adding or modifying the 
parameters. Specifically, attackers can introduce a backdoor task while ensuring normal training of the main task. The main task aims to recover the original transmitted image under channel conditions characterized by $\epsilon$, whereas the backdoor task is designed to reconstruct a target image $x_a$ under a specific channel parameters $\epsilon_{target} = \mathcal{T}(\epsilon) $, which exist in the practical environment. Under normal circumstances, the SemCom system operates normally in an environment with channel characteristics of $\epsilon$. However, when the environment changes and the channel characteristics become $\epsilon_{target}$, the backdoor will automatically be triggered. The high dimensionality of the wireless parameter space provides immense flexibility for trigger selection.
Within this framework, we propose two distinct backdoor trigger construction strategies:

1) $n$-trigger: The adversary selects a specific noise standard deviation $\sigma_a$ as the activation condition, defined as $\mathcal{T}(\epsilon)=\{\sigma_a\}$. The backdoor is triggered when the channel noise power reaches $\sigma_a^2$, inducing the decoder to output the target image $x_a$.

2) $H$-trigger: The adversary utilizes the channel gain $H$ as the trigger, defined by the parameter set $\mathcal{T}(\epsilon)=\{\sigma_n, \sigma_h\}$. For example, by targeting a Rayleigh fading distribution where $H \sim \mathcal{CN}(0, \sigma_h^2 I_k)$, the adversary ensures the backdoor is activated only under specific fading intensities typical of certain geographic or environmental contexts.

\begin{algorithm}[t]
\DontPrintSemicolon
\SetAlgoLined
\caption{Backdoor Training Process.}
\label{alg:CTBA}
\KwIn{Batch input $\mathbf{X} \in \mathbb{R}^{B \times h \times w \times c}$, Batch size $B$, target image $\mathbf{x}_a$, poison ratio $P$, encoder $f_\theta$, decoder $g_\phi$, channel models $\eta$}

Calculate poison number $N_p=BP$\;
\For{every epoch}{
    \For{every batch}{
    Batch input $\mathbf{X} \in \mathbb{R}^{B \times h \times w \times c}$\;
    $\mathbf{Z}\leftarrow f_\theta(\mathbf{X})$\;
    $(\mathbf{Z}_b, \mathbf{Z}_c) \leftarrow (\mathbf{Z}[0:N_p], \mathbf{Z}[N_p:\mathrm{end}])$ \;
    $\hat{\mathbf{Z}}_b ,\hat{\mathbf{Z}}_c\leftarrow \eta(\mathbf{Z}_b;\mathcal{T}(\epsilon)),\eta(\mathbf{Z}_c;\epsilon) $\;
    $\mathbf{Y} \leftarrow \mathrm{concat}(\mathbf{x_a},\mathbf{X}[N_p:\mathrm{end}] )$\;
    Calculate $\mathcal{L}(\hat{\mathbf{X}},\mathbf{Y})$\;
    Update Parameter $(\theta,\phi)$\;
    }
}
\KwOut{Optimized parameters $(\theta^*, \phi^*)$}
\end{algorithm}

\subsection{Backdoor training}

With the above analysis, the training objective for backdoor models is to simultaneously minimize the average loss of the main task and the average loss of the backdoor task. Our CT-BA method adopts the conventional label-poisoning backdoor training procedure but modifies it to suit the channel-triggered attack scenario.

As illustrated in Fig. \ref{fig_4}, let $\mathbf{X} = \{x_k\}_{k=1}^N$ represent the input dataset, where $N$ is the total number of training samples. A subset of $\mathbf{X}$, denoted as $\mathbf{X}_a$, is replaced with target images $x_a$ to construct the poisoned label set $\mathbf{Y} = \mathbf{X}_c \cup \mathbf{X}_a$, where $|\mathbf{X}_a| = NP$, and $P$ represents the poisoning ratio. The encoder $f_\theta$ maps each input sample $x_k$ to a channel symbol $z_k = f_\theta(x_k)$. Consequently, the symbol set for the entire training set is denoted as $\mathbf{Z} = \{z_k\}_{k=1}^N$. 

To implement the backdoor attack, we sample a subset $\mathbf{Z}_b$ of $\mathbf{Z}$ as the to-be-poisoned symbol subset, where $|\mathbf{Z}_b| = PN$, and $P$ represents the poisoning ratio. The remaining symbols form the clean symbol subset $\mathbf{Z}_c$, such that $\mathbf{Z} = \mathbf{Z}_c \cup \mathbf{Z}_b$. 

To accomplish trigger injection, the to-be-poisoned symbols are passed through the channel with trigger $\eta(\cdot; \mathcal{T}(\epsilon))$, generating the set of poisoned symbols $\hat{\mathbf{Z}}_b = \{\eta(z; \mathcal{T}(\epsilon)) \mid z \in \mathbf{Z}_b\}$. For reconstructing the target image $x_a$ from the poisoned symbols, we minimize the following loss function for the backdoor task:

\begin{equation}
    \mathcal{L}_\text{adv}(\theta, \phi) = \sum_{\hat{z} \in \hat{\mathbf{Z}}_b} \mathcal{L}(g_\phi(\eta(z; \mathcal{T}(\epsilon ))), x_a).
\end{equation}

Conversely, the clean symbol subset $\mathbf{Z}_c$ passes through a clean channel, resulting in $\hat{\mathbf{Z}}_c = \{\eta(z; \epsilon) \mid z \in \mathbf{Z}_c\}$. We define the clean dataset as 
$\mathcal{D}_c = \{(x, \hat{z}) \mid \hat{z} = \eta(f_\theta(x); \epsilon), x \in {\mathbf{X}}_c \}$.
To ensure effective reconstruction of clean outputs from clean symbols, we optimize the following loss function for the primary task:

\begin{equation}
    \mathcal{L}_\text{clean}(\theta, \phi) = \sum_{(x,\hat z) \in \mathcal{D}_c }\mathcal{L}(g_\phi(\eta(z;\epsilon), x).
\end{equation}

The overall objective can be formalized as the following optimization problem:

\begin{equation}
    \min_{\theta, \phi} \mathcal{L}_\text{total} = \alpha \mathcal{L}_\text{adv}(\theta, \phi)+(1-\alpha) \mathcal{L}_\text{clean}(\theta, \phi),
\end{equation}
where $\alpha$ controls the mixing strength of the losses. We perform poisoning in each training batch to enhance the performance of the backdoor task. Setting $\alpha$ to $P$ allows the mixing strength of the losses to be controlled by the poisoning ratio, such that $\mathcal{L}_\text{total} = \mathcal{L}(\hat{\mathbf{X}}, \mathbf{Y})$. 
The complete backdoor attack procedure is formally described in Algorithm \ref{alg:CTBA}, which systematically outlines the key stages of the malicious operation.

\subsection{Practical Feasibility and robustness analysis}
Based on the above analysis, we have presented two different design ideas for the backdoor trigger. The attacker can pre-embed different triggers during the backdoor training stage to achieve different attack effects in the actual inference stage according to their intentions.

During the reasoning stage, on one hand, the attacker can trigger the backdoor automatically without participating in any activation process. For instance, dense urban environments naturally induce rich multipath scattering, which is characterized by Rayleigh fading. An adversary can train a backdoor (H-trigger) specifically activated by this fading distribution. Consequently, when an enemy drone or autonomous vehicle transitions from a line-of-sight environment into a dense urban area, the change in channel statistics automatically activates the backdoor—potentially causing the reconstruction of misleading visual data (e.g., removing obstacles) and leading to a crash or mission failure. On the other hand, if immediate activation is required, the adversary can employ an active strategy using the noise-based trigger (n-trigger). For instance, The attacker can pre-embed a backdoor that is triggered under a low signal-to-noise ratio condition during the training phase. As the attacker is involved in the training process, they gain prior knowledge of the semantic communication model's transmission power. Furthermore, the attacker can readily ascertain the statistical properties of the target wireless channel's environmental noise because this information is typically public knowledge or easily measured, a standard capability assumed for  adversaries in wireless environments. They can simply employ a jammer to inject Gaussian noise, raising the noise floor to the specific power level ($\sigma_a$) required to activate the backdoor. 

It should be noted that in actual communication systems, there are safeguards such as adaptive modulation and error correction. We acknowledge that incorporating such protection measures into SemCom systems is technically feasible. However, the fundamental E2E characteristic of SemCom requires that any safeguard module introduced must be  differentiable to ensure effective backpropagation of the gradient during training. Nonetheless, the CT-BA is still able to bypass these protection mechanisms. Our CT-BA is implemented through an E2E joint optimization paradigm, in which the semantic encoder and decoder are jointly optimized during the training process containing label poisoning to simultaneously minimize the loss functions of the main task and the backdoor task. Since the entire communication chain is implemented as differentiable layers, the gradient of the backdoor task can be propagated through the complex physical layer operations using the chain rule during the backpropagation process. This enables the semantic decoder to internalize the statistical patterns of the triggers as the intrinsic semantic features of its optimization objective, rather than merely treating them as noise that needs to be filtered.

\section{Simulation Results}

\begin{table*}[htbp]
\centering
\caption{Attack performance on different datasets and bandwidth ratios. The backdoor task is trained and tested at SNR=-15 dB, while the main task is trained and tested at SNR=15 dB.}
\label{tab:Dataset_Bandwidth_performance}
\begin{tabular}{@{}cc|cc|cc|c@{}}
\toprule
\multicolumn{2}{c|}{Aspect $\rightarrow$} & \multicolumn{2}{c|}{Effectiveness} & \multicolumn{2}{c|}{Stealthiness} & Robustness \\ \midrule
Datasets$\downarrow$ & R$\downarrow$ & $PSNR_b$ (dB) & ASR (\%) & $PSNR_m / PSNR_c$ (dB) & AEVC / CA (\%) & AC (\%) \\ \midrule
\multirow{3}{*}{CIFAR-10} & 1/4 & 57.04$\pm$0.11 & \textbf{100.0$\pm$0.00} & 46.65$\pm$0.30 / 46.36$\pm$0.25 & 95.12$\pm$1.70 / 95.17$\pm$1.83 & 9.65$\pm$2.48 \\
 & 1/6 & 51.35$\pm$0.06 & \textbf{100.0$\pm$0.00} & 44.27$\pm$0.29 / 41.97$\pm$0.28 & 95.09$\pm$1.90 / 94.91$\pm$1.82 & 9.67$\pm$2.49 \\
 & 1/12 & 46.22$\pm$0.01 & \textbf{100.0$\pm$0.00} & 42.37$\pm$0.24 / 39.30$\pm$0.25 & 94.74$\pm$2.02 / 94.81$\pm$1.91 & 9.68$\pm$2.48 \\ \midrule
\multirow{3}{*}{MNIST} & 1/4 & 61.47$\pm$0.13 & \textbf{100.0$\pm$0.00} & 54.32$\pm$0.31 / 55.79$\pm$0.20 & 99.70$\pm$0.53 / 99.71$\pm$0.53 & 9.55$\pm$1.93 \\
 & 1/6 & 61.40$\pm$0.10 & \textbf{100.0$\pm$0.00} & 54.07$\pm$0.29 / 54.83$\pm$0.24 & 99.70$\pm$0.56 / 99.70$\pm$0.56 & 9.38$\pm$2.11 \\
 & 1/12 & 55.91$\pm$0.02 & \textbf{100.0$\pm$0.00} & 50.03$\pm$0.30 / 52.56$\pm$0.21 & 99.69$\pm$0.51 / 99.69$\pm$0.53 & 9.38$\pm$1.93 \\ \midrule
\multirow{3}{*}{ImageNet} & 1/4 & 34.24$\pm$3.37 & \textbf{99.82$\pm$0.54} & 32.06$\pm$1.20 / 33.80$\pm$1.18 & 79.63$\pm$10.01 / 80.36$\pm$9.93 & 0.10$\pm$1.37 \\
 & 1/6 & 32.81$\pm$2.97 & \textbf{99.76$\pm$0.57} & 30.15$\pm$1.18 / 31.20$\pm$1.18 & 78.41$\pm$10.34 / 79.25$\pm$10.11 & 0.90$\pm$1.36 \\
 & 1/12 & 26.73$\pm$2.20 & \textbf{99.24$\pm$0.99} & 27.05$\pm$1.15 / 27.65$\pm$1.16 & 74.25$\pm$11.26 / 75.26$\pm$11.07 & 0.09$\pm$1.29 \\ \bottomrule
\end{tabular}
\end{table*}

\begin{table*}[t]
\centering
\caption{Attack performance on different poision Ratio and main task SNR. The backdoor task trained and tested at SNR=-10 dB}
\label{tab:PoisioinRatio_SNR_Performance}
\begin{tabular}{@{}cc|cc|cc|c@{}}
\toprule
\multicolumn{2}{c|}{Aspect$\rightarrow$} & \multicolumn{2}{c|}{Effectiveness} & \multicolumn{2}{c|}{Stealthiness} & Robustness \\ \midrule
P$\downarrow$ & $\mathrm{SNR_{train}}$(dB)$\downarrow$ & $\mathrm{PSNR_b}$(dB) & ASR(\%) & $\mathrm{PSNR_m}$/$\mathrm{PSNR_c}$(dB) & AEVC(\%) & AC(\%) \\ \midrule
\multirow{3}{*}{1\%} & 1 & 42.30$\pm$3.80 & \textbf{99.90$\pm$0.09} & 42.09$\pm$0.25 / 41.83$\pm$0.24 & 94.80$\pm$1.81 / 95.13$\pm$1.98 & 9.63$\pm$2.47 \\
 & 5 & 43.36$\pm$0.72 & \textbf{100.0$\pm$0.00} & 43.33$\pm$0.26 / 42.96$\pm$0.23 & 94.88$\pm$1.80 / 94.79$\pm$1.99 & 9.61$\pm$2.46 \\
 & 10 & 43.58$\pm$0.36 & \textbf{100.0$\pm$0.00} & 44.78$\pm$0.28 / 43.79$\pm$0.25 & 94.95$\pm$1.86 / 94.9$\pm$1.83 & 9.58$\pm$2.45 \\ \midrule
\multirow{3}{*}{5\%} & 1 & 48.58$\pm$4.25 & \textbf{100.0$\pm$0.00} & 41.59$\pm$0.26 / 41.83$\pm$0.24 & 94.80$\pm$1.84 / 94.7$\pm$2.01 & 9.62$\pm$2.51 \\
 & 5 & 49.56$\pm$0.38 & \textbf{100.0$\pm$0.00} & 43.12$\pm$0.26 / 42.96$\pm$0.23 & 94.88$\pm$1.92 / 95.03$\pm$2.00 & 9.49$\pm$2.52 \\
 & 10 & 49.47$\pm$0.35 & \textbf{100.0$\pm$0.00} & 44.54$\pm$0.27 / 43.78$\pm$0.25 & 94.95$\pm$1.86 / 94.9$\pm$1.83 & 9.57$\pm$2.51 \\ \midrule
\multirow{3}{*}{10\%} & 1 & 50.38$\pm$0.84 & \textbf{100.0$\pm$0.00} & 41.85$\pm$0.25 / 42.01$\pm$0.24 & 94.81$\pm$1.91 / 94.74$\pm$2.03 & 9.61$\pm$2.54 \\
 & 5 & 51.00$\pm$0.34 & \textbf{100.0$\pm$0.00} & 43.72$\pm$0.25 / 42.96$\pm$0.23 & 94.88$\pm$1.83 / 94.88$\pm$1.99 & 9.55$\pm$2.53 \\
 & 10 & 50.83$\pm$0.02 & \textbf{100.0$\pm$0.00} & 45.82$\pm$0.26 / 43.79$\pm$0.25 & 94.95$\pm$1.95 / 95.02$\pm$1.83 & 9.60$\pm$2.45 \\ \bottomrule
\end{tabular}
\end{table*}

\subsection{Simulation Setup}
\subsubsection{Simulation Environment}In our experiments, all models are implemented using PyTorch 2.4.0 and trained on two NVIDIA A40 GPUs. We employ the Adam optimizer for model optimization. For the ViT-based JSCC model, the patch size ($p$) is set to 16, while the parameters $L_t$ and $L_d$ are set to 32 and 8, respectively. Each Transformer layer consists of $N_s = 16$ self-attention heads. Training continues until validation performance plateaus. To ensure stable convergence, we apply a learning rate warmup strategy for the first 40 epochs, followed by learning rate decay. The initial learning rate is determined as $0.001 \times \text{total\_batch\_size} / 256$, with a batch size of 128.
\subsubsection{Dataset, model, and baseline}We evaluate CT-BA on the MNIST \cite{lecun1998gradient}, CIFAR-10 \cite{krizhevsky2009learning}, and ImageNet \cite{deng2009imagenet} datasets. The MNIST dataset consists of 70,000 square ($28 \times 28 = 784$  pixel) grayscale handwritten digital images, divided into 10 categories (60,000 for training and 10,000 for testing). The CIFAR-10 dataset consists of 60,000 square ($3 \times 32 \times 32 = 3072$ pixel) 3-channel color images, divided into 10 categories (50,000 for training and 10,000 for testing). The ImageNet dataset contains over 14 million high-resolution color images, categorized into 1,000 classes. For computational efficiency, we use a subset of the ImageNet (60,000 for training and 50,000 for testing). To assess robustness, we test CT-BA on a ViT-based JSCC communication system. Besides, we evaluate the universality of our attack on three E2E SemCom systems: BDJSCC, ADJSCC,  JSCCOFDM, and VIT-MIMO. 
For each backdoor model, we train a corresponding benign model as the baseline. These benign models are trained with the same hyperparameters (such as learning rate, number of training epochs) as the backdoor models, except that the benign model is trained end-to-end on the channel characterized by $\epsilon$ to restore the image capability (the main task), while the backdoor model is trained end-to-end on the channel characterized by $\mathcal{T}(\epsilon)$ to restore the target image capability (the backdoor task) simultaneously during the training of the primary task.  These benign models serve as a reference to measure backdoor attack impact.

\subsubsection{Evaluation Metrics}To evaluate the attack performance, seven key metrics are employed: benign model reconstruction PSNR ($\mathrm{PSNR_c}$), main task reconstruction PSNR ($\mathrm{PSNR_m}$), backdoor task reconstruction PSNR ($\mathrm{PSNR_b}$), clean accuracy (CA), accuracy excluding victim class (AEVC), attack success rate (ASR), and average confusion (AC). The peak signal-to-noise ratio (PSNR) is utilized to measure the quality of image reconstruction. Specifically, $\mathrm{PSNR_b}$ represents the PSNR of the backdoored model in reconstructing target images for the backdoor task, while $\mathrm{PSNR_m}$ denotes the PSNR for the main task. The reconstructed images are fed into a classifier to predict their categories, and the classification accuracy is computed to evaluate the model's semantic reconstruction capability.

Considering  $\Omega$ as the classification model, the backdoored model is denoted as $\Phi_b$ and  the benign model is denoted as $\Phi_c$. The definitions of accuracy-related metrics are as follows,
\allowdisplaybreaks 
\begin{subequations}
\label{eq:main}
\begin{align}
CA &= \frac{\sum_{i=1}^{N} \delta [\Omega (\Phi_{c}(x_{i})), y_{i}]}{N} \label{eq:CA} \\
AEVC &= \frac{\sum_{i=1}^{N} \delta [\Omega (\Phi_{b}(x_{i})), y_{i}]}{N} \label{eq:AEVC} \\
ASR &= \frac{\sum_{i=1}^{N} \delta [\Omega (\Phi_{b}(A(x_{i}))), y_{t}]}{N} \label{eq:ASR} \\
AC &= \frac{\sum_{i=1}^{N} \delta [\Omega (\Phi_{b}(x_{i})), y_{t}]}{2N} \nonumber \\
&\quad + \frac{\sum_{i=1}^{N} \delta [\Omega (\Phi_{b}(A(x_{i}))), y_{i}]}{2N} \label{eq:AC}
\end{align}
\end{subequations}
where $\delta [a,b] = 1$ if $a \ne b$ and $\delta [a,b] = 0$ otherwise. Here, $A(\cdot)$ denotes the backdoor trigger operation, $y_i$ represents the classification labels, and $y_t$ is the target label. A higher $\mathrm{PSNR_b}$ indicates superior performance in the backdoor task. The closer AEVC is to CA, the better the backdoor can be concealed, as the backdoored model behaves similarly to the benign model for clean inputs. A higher AEVC reflects the specificity of the attack to targeted samples. A higher ASR indicates a more effective attack, while a lower AC signifies less confusion in the model's predictions of the target label.

\subsection{Backdoor Attack Performance Evaluation} 

    
    
    

\begin{figure*}[t] 
  \centering
  
  \begin{minipage}{0.48\textwidth}
    \centering
    \includegraphics[width=\linewidth]{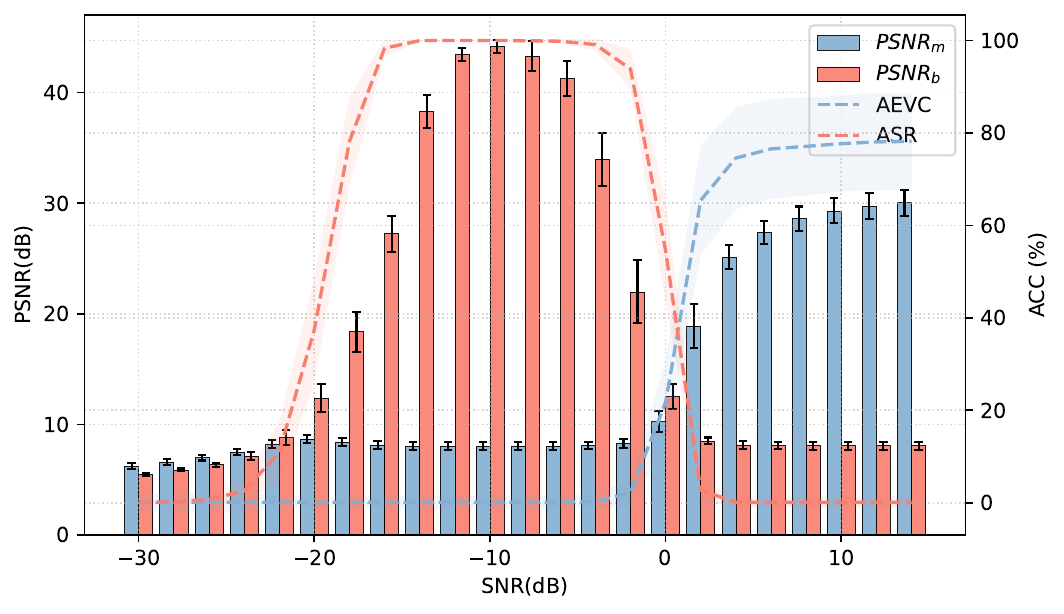}
    \caption{Performance of the n-triggered backdoored model across varying test SNRs on the ImageNet dataset.}
    \label{fig_7}
  \end{minipage}
  \hfill 
  \begin{minipage}{0.48\textwidth}
    \centering
    \includegraphics[width=\linewidth]{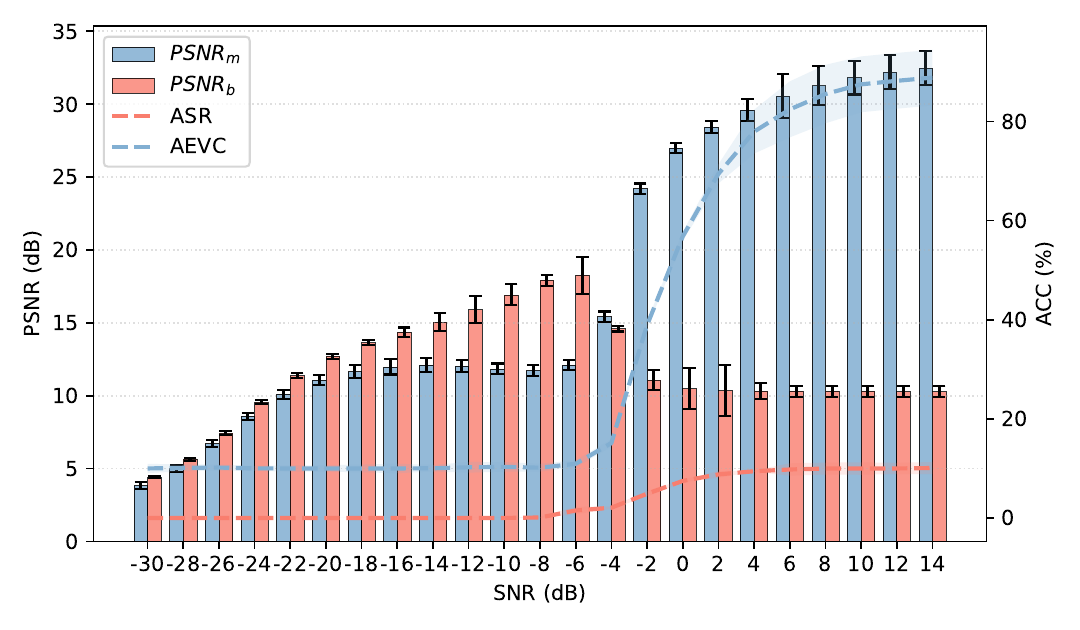}
    \caption{Performance of the n-triggered backdoored model across varying test SNRs on the ImageNet dataset.}
    \label{fig_6}
  \end{minipage}

  \vspace{2em} 

  \begin{minipage}{0.48\textwidth}
    \centering
    \includegraphics[width=\linewidth]{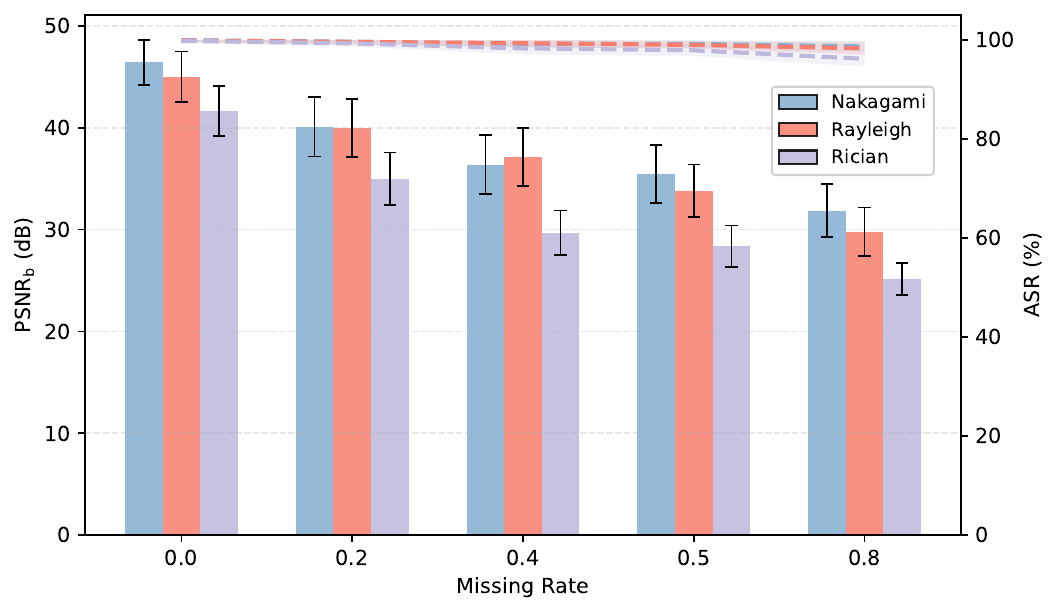}
    \caption{Impact of channel state missing rate on both reconstruction PSNR and attack success rate (ASR) across diverse fading environments.}
    \label{fig:partial-csi}
  \end{minipage}
  \hfill
  \begin{minipage}{0.48\textwidth}
    \centering
    \includegraphics[width=\linewidth]{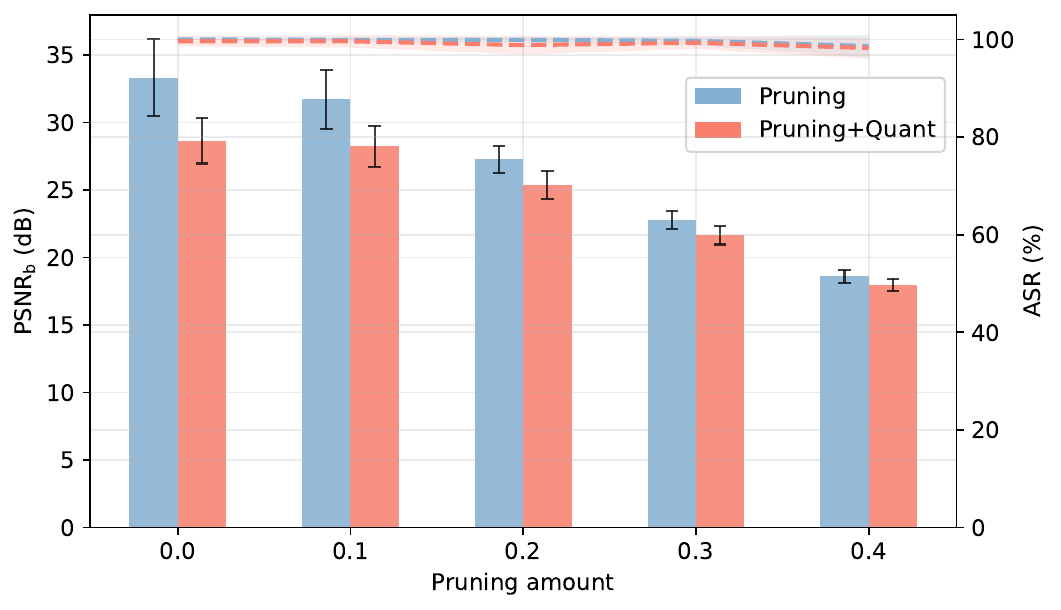}
    \caption{Impact of model pruning (L1Unstructured) and quantization (INT8) on CT-BA performance.}
    \label{fig:pruning}
  \end{minipage}

\end{figure*}






\subsubsection{Robustness Across Datasets and Model Bandwidth Ratio} 

\begin{figure*}[!t]
\centering
\includegraphics[width=0.9\textwidth]{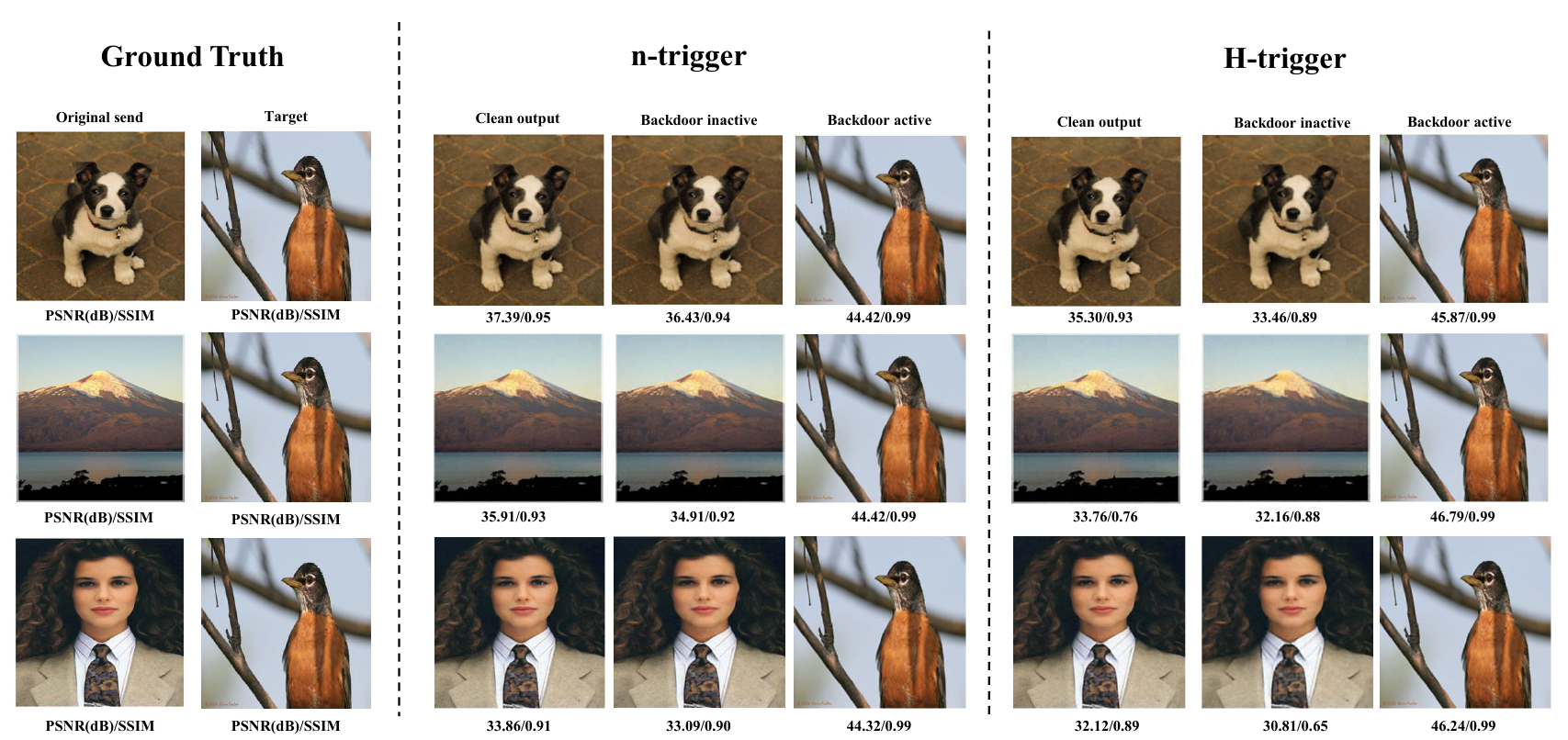}
\caption{Visualization of backdoor model performance under two different trigger strategies, trained on high-quality datasets using a ViT-based JSCC architecture with $R = 1/6$ and $P = 0.05$.}
\label{fig_5}
\end{figure*}

\begin{table*}[htbp]
\centering
\caption{Attack perform under diferent channel conditions. The model is trained and tested on AWGN channels at SNR=15 dB.}
\label{tab:tab_3}
\begin{tabular}{@{}l|ll|ll|l@{}}
\toprule
Aspect$\rightarrow$ & \multicolumn{2}{c|}{Effectiveness} & \multicolumn{2}{c|}{Stealthiness} & \multicolumn{1}{c}{Robustness} \\ \midrule
Channel$\downarrow$ & \multicolumn{1}{c}{$\mathrm{PSNR_b}$(dB)} & \multicolumn{1}{c|}{ASR(\%)} & \multicolumn{1}{c}{$\mathrm{PSNR_m}$(dB)/$\mathrm{PSNR_c}$(dB)} & \multicolumn{1}{c|}{AEVC(\%)/CA(\%)} & \multicolumn{1}{c}{AC(\%)} \\ \midrule
Rayleigh & 45.79$\pm$4.81 & 99.89$\pm$0.27 & 39.28$\pm$0.26/43.26$\pm$0.24 & 94.75$\pm$1.97/94.88$\pm$2.02 & 9.97$\pm$2.73 \\
Nakagami-m & 46.57$\pm$4.49 & 99.93$\pm$0.24 & 39.85$\pm$0.27/43.26$\pm$0.24 & 94.81$\pm$1.93/94.88$\pm$2.02 & 10.0$\pm$2.68 \\
Rician & 40.27$\pm$6.82 & 99.66$\pm$0.52 & 37.39$\pm$0.73/43.26$\pm$0.24 & 94.28$\pm$1.98/94.88$\pm$2.02 & 10.2$\pm$2.66 \\ \bottomrule
\end{tabular}
\end{table*}


This experiment evaluate the robustness of the CT-BA across different datasets and model bandwidth ratios over an AWGN channel. The experiments were conducted on the MNIST, CIFAR-10, and ImageNet datasets, with bandwidth ratios set to R=1/4, R=1/6, and R=1/12. For the backdoored model, the main task was trained and tested at an SNR of 15 dB, while the backdoor task was trained and tested at an SNR of -15 dB. The benign model was trained and tested exclusively on the main task at an SNR of 15 dB for comparison. 


As shown in Table \ref{tab:Dataset_Bandwidth_performance}, the main task reconstruction quality ($\mathrm{PSNR_m}$) closely aligns with the benign model performance ($\mathrm{PSNR_c}$). Nevertheless, the semantic difference between them is negligible, which indicates that the backdoor attack effectively maintains the SemCom quality of the normal task while introducing no detectable semantic confusion due to the attack functionality. 

Furthermore, the backdoor task achieves higher reconstruction quality in all tests. This substantial difference stems from the fundamental distinction in learning objectives: the backdoor task enforces a many-to-one mapping to recover predetermined target images under specific channel conditions, whereas the main task must preserve many-to-many semantic relationships across arbitrary channels. 

Near-identical CA and  AEVC, where the maximum discrepancy is $\Delta\mathrm{CA-AEVC}\leq1.01\%$ (ImageNet, R=1/12), confirms the backdoor’s imperceptibility during normal operation. The high AEVC values reveal that the backdoor neither degrades normal SemCom nor introduces detectable anomalies. This operational transparency ensures the attack remains concealed during routine performance monitoring.

The ASR=100.0\% across all test cases demonstrates the deterministic activation of backdoor behavior under targeted channel conditions. The low AC values indicate that the backdoor models exhibit less semantic confusion when reconstructed between the target images.

\subsubsection{Sensitivity to SNR and Poisoning Ratios Variations} 
This experiment evaluates the sensitivity of CT-BA to variations in training SNR and \textit{Poisoning Ratios} over AWGN channels. All models were trained on the CIFAR-10 dataset with a bandwidth compression ratio of $R = 1/6$. The backdoor models were trained with poisoning rates ranging from $P = 1\%$ to $P = 10\%$, while the main task training SNR ($\mathrm{SNR_{train}}$) was set to $\mathrm{1~dB}$, $\mathrm{5~dB}$, and $\mathrm{10~dB}$. The backdoor task was consistently trained at $\mathrm{-10~dB}$ SNR. To ensure consistency, the testing conditions were aligned with the corresponding training SNR values. Additionally, benign models were trained as baselines for comparison, with $\mathrm{SNR_{train}}$ values of $\mathrm{1~dB}$, $\mathrm{5~dB}$, and $\mathrm{10~dB}$.

As shown in Table \ref{tab:PoisioinRatio_SNR_Performance}, increasing the \textit{Poisoning Ratios} under fixed training SNR conditions (e.g., $\mathrm{SNR_{train}} = 1~\mathrm{dB}$) effectively amplifies backdoor reconstruction fidelity while preserving stealthiness (evidenced by the close alignment between $\mathrm{PSNR_m}$ and $\mathrm{PSNR_c}$ and minimal discrepancies between CA and AEVC).  Notably, even minimal \textit{Poisoning Ratios} (e.g., $P=1\%$) achieve robust backdoor performance with $\mathrm{PSNR_b}$ exceeding $\mathrm{PSNR_m}$ by $17.92~\mathrm{dB}$, while maintaining ASR values above $99.99\%$. For a fixed \textit{Poisoning Ratios}, elevating the $\mathrm{SNR_{train}}$ from $\mathrm{1~dB}$ to $\mathrm{10~dB}$ the main task performance is improved with the improvement of $\mathrm{SNR_{train}}$, while the backdoor performance remains stable (ASR=100.0\%,$\Delta\mathrm{CA-AEVC}\leq0.05\%$). 

Furthermore, we evaluate models across test SNR ranging from $\mathrm{-30~dB}$ to $\mathrm{15~dB}$ to observe the main task performance and backdoor performance. As illustrated in Fig.\ref{fig_7},  the backdoor task achieves the best performance on the predefined backdoor trigger SNR while retaining functional activation within a certain range,  but this does not affect the performance of the main task on its predetermined training SNR. 

\subsubsection{Trigger mechanism specificity} 



To further validate the channel-specific activation of the backdoor, we analyze the performance of the backdoor model under varying SNR conditions in AWGN channels. As shown in Fig.~\ref{fig_6}, the $\mathrm{PSNR_m}$ increases as the SNR rises. Correspondingly, the  AEVC improves from 9.9\% to 88.86\% across the same SNR range. In contrast, the  $\mathrm{PSNR_b}$ remains below 15 dB for most SNR values, peaking at 18.24 dB only at -6 dB SNR. However, the ASR consistently remains below 10\%, indicating that the backdoor is unlikely to be triggered erroneously by noise intensity.

The monotonic improvement in $\mathrm{PSNR_m}$ with increasing SNR aligns with expected communication system behavior, where higher SNR enhances signal recovery and sematic accuracy. However, the persistently low $\mathrm{PSNR_{b}}$ and ASR values across all SNR levels demonstrate that the backdoor remains dormant in AWGN channels, regardless of noise conditions. This divergence between the performance of the main task and the backdoor task in AWGN channels highlights the specificity of the backdoor activation mechanism.

\subsubsection{Attack perform under more complex channel conditions} 
To validate the robustness and practical applicability of the proposed framework, we extended our evaluation beyond the idealized AWGN channel to more complex fading environments and impaired CSI scenarios. As presented in Tab. \ref{tab:tab_3}, we evaluated the attack effectiveness across Rayleigh, Nakagami-$m$ ($m=0.6, \Omega=1$), and Rician ($K=5$) fading channels using the ViT-based JSCC architecture on the CIFAR-10 dataset. In terms of effectiveness, the proposed CT-BA demonstrates high performance across all evaluated datasets and bandwidth ratios, achieving a near-perfect attack success rate and high-fidelity target reconstruction. Regarding stealthiness, the attack exhibits excellent concealment as the main task reconstruction quality ($PSNR_m$) closely aligns with the benign model performance ($PSNR_c$), and the negligible gap between AEVC and CA ensures no detectable semantic confusion during normal operation. These results collectively reflect the robust capability of the attack to maintain deterministic activation and operational transparency across diverse channel conditions and communication constraints. We further investigate the performance of the proposed backdoor under partial CSI by varying the channel state missing rate across Rayleigh, Nakagami, and Rician fading environments, as illustrated in Fig. \ref{fig:partial-csi}. The experimental results, presented via a dual-axis representation, show that while the reconstruction PSNR naturally declines with increasing information loss, the attack success rate (ASR) remains remarkably stable and high across all tested scenarios, concluding that the CT-BA mechanism possesses robustness against imperfect channel state information in realistic communication settings.

\begin{table*}[t]
\centering
\caption{The comparison of different domain trigger on the dpjscc model.The trigger in the channel domain demonstrated greater effectiveness and stealthinessacross all the test data.}
\label{tab:diferentDomain}
\begin{tabular}{@{}c|cc|cc|c@{}}
\toprule
Aspect$\rightarrow$ & \multicolumn{2}{c|}{Effectiveness} & \multicolumn{2}{c|}{Stealthiness} & Robustness \\ \midrule
strategy$\downarrow$ & $\mathrm{PSNR_b}$(dB) & ASR(\%) & $\mathrm{PSNR_m}$(dB)/$\mathrm{PSNR_c}$(dB) & AEVC(\%)/CA(\%) & AC(\%) \\ \midrule
BadNets & 20.28$\pm$1.49 & 93.31$\pm$1.98 & 33.92$\pm$0.25 / 34.42$\pm$0.27 & 93.26$\pm$2.18 / 93.4$\pm$1.89 & 11.7$\pm$2.47 \\
WaNets & 20.21$\pm$1.49 & 93.32$\pm$1.98 & 33.92$\pm$0.25 / 34.42$\pm$0.27 & 93.35$\pm$2.18 / 93.4$\pm$1.89 & 11.8$\pm$2.45 \\
Ours & 34.09$\pm$1.28 & 100$\pm$0.00 & 34.52$\pm$0.24 / 34.42$\pm$0.27 & 93.6$\pm$1.98 / 93.4$\pm$1.89 & 9.94$\pm$2.26 \\ \bottomrule
\end{tabular}
\end{table*}

\begin{table}[!t]
\caption{Attack performance on typical systems. For BDJSCC and JSCCOFDM, the backdoor tasks are trained and tested at SNR=0 dB, while the main task is trained and tested at SNR=20 dB. For ADJSCC, the backdoor tasks are trained and tested at SNR=-2 dB, whereas the main task is trained across SNR=[0,20] dB and tested at SNR=20 dB.}
\label{tab:other_system_perfomence}
\begin{tabular}{ccccc}
\hline
Model & BDJSCC & ADJSCC & JSCCOFDM & VIT-MIMO \\ \hline
$\mathrm{PSNR}_{c}$(dB) & 34.43 & 36.47 & 31.97 & 31.40 \\
$\mathrm{PSNR}_{m}$(dB) & 34.53 & 36.45 & 31.85 & 30.16 \\
$\mathrm{PSNR}_{b}$(dB) & 34.09 & 55.74 & 36.94 & 45.65 \\
CA(\%) & 93.41 & 94.07 & 91.67 & 92.45 \\
AEVC(\%) & 93.61 & 93.96 & 91.54 & 92.12 \\
ASR(\%) & 100.0 & 100.0 & 100.0 & 100.0 \\
AC(\%) & 9.940 & 9.960 & 9.560 & 9.230 \\ \hline
\end{tabular}
\end{table}


\subsubsection{Common deployment constraints} 
To evaluate the persistence of CT-BA during model optimization for edge deployment, we investigated the impact of model pruning (L1 Unstructured) and quantization (INT8) on the ViT-based JSCC architecture using the ImageNet dataset. As illustrated in Fig. \ref{fig:pruning}, which utilizes a dual-axis representation to show $\mathrm{PSNR_b} $ and ASR simultaneously, the proposed attack exhibits remarkable stability against model compression. While the backdoor task $\mathrm{PSNR_b}$ predictably decreases as the pruning amount increases from 0.0 to 0.4, the ASR remains consistently near 100\% even under the joint application of pruning and quantization. These results lead to the conclusion that the channel-triggered backdoor mapping is deeply embedded within the model's fundamental feature representations, making it resistant to standard model compression techniques.

\subsubsection{Comparing to input-triggered attacks} 
To further illustrate the stealthiness and effectiveness of CT-BA, we conducted an experiment and compared it with the put-triggered attacks. In order to conduct a unified comparison, we applied the WaNet trigger construction pattern to the semantic communication system. As shown in Tab. \ref{tab:diferentDomain}, we compared the effectiveness and stealthiness  of channel-triggered attacks and input-triggered attacks on the BDJSCC model. To ensure fairness, the training of the experimental model used the same parameters, with a poisoning rate set at 0.01 and a bandwidth compression ratio of 1/6 for BDJSCC. Additionally, the transmission power of BDJSCC was 1(w), the channel trigger was set to $\sigma_a=1$ (with a SNR of 0 dB), and the main task run at a SNR of 20 dB. In terms of effectiveness, the target image reconstruction PSNR ($PSNR_b$) and attack success rate (ASR) are both superior to the input-triggered attack. In terms of stealthiness, The PSNR ($PSNR_m$) of the backdoor model when reconstructing clean input is closer to the reconstruction performance of the benign model ($PSNR_c$). Accordingly, the clean accuracy (CA) and the accuracy excluding the victim class (AEVC) are also closer. Therefore, channel-triggered attacks outperform input-triggered attacks in both effectiveness and concealment. This is because, compared to the traditional input-based attack, the proposed CT-BA utilizes global channel statistics to inherently interfere with each dimension of the transmitted symbol sequence, and the semantic communication transmits semantic information, with the trigger directly acting on the feature domain. This global footprint makes the trigger feature significantly more salient and distinguishable for the decoder.

\subsubsection{High Resolution Dataset and Visualization}
To further validate the effectiveness and channel-specific behavior of the proposed backdoor mechanism, we conducted a visualization  on a high-quality dataset. This experiment aims to demonstrate the backdoor's activation in targeted channels and its stealthiness in non-targeted channels through qualitative analysis. Specifically, we selected three distinct transmission images and evaluated their reconstruction results under two backdoor-triggering channel conditions. We validate the CT-BA on the ImageNet dataset using R=1/6. For $n$-trigger, backdoor task testing SNR in $\mathrm{-15~dB}$ and main task testing SNR in $\mathrm{15~dB}$. For $H$-trigger, backdoor task testing in Rayleigh channel with testing SNR $\mathrm{15~dB}$ and main task testing in AWGN channel with testing SNR $\mathrm{15~dB}$.

As shown in Fig.~\ref{fig_5}, for each test sample, the system successfully reconstructs target images with high fidelity when the transmitted symbols satisfy the predefined trigger conditions. Conversely, when the backdoor remains inactive, the backdoored model maintains nearly identical visual performance to its clean counterpart, with imperceptible degradation in output quality. The visualization results demonstrate that the clean reconstruction for the benign model and the normal reconstruction for the backdoored model under both $n$-trigger and $H$-trigger schemes are nearly indistinguishable, with no perceptible differences to the human eye. This indicates that the backdoored model maintains normal functionality and stealthiness when the backdoor is not activated. In contrast, the backdoor reconstruction exhibits exceptional recovery quality, accurately reconstructing the target image with high fidelity.

\begin{figure*}[t]
\centering
\includegraphics[width=0.70\textwidth]{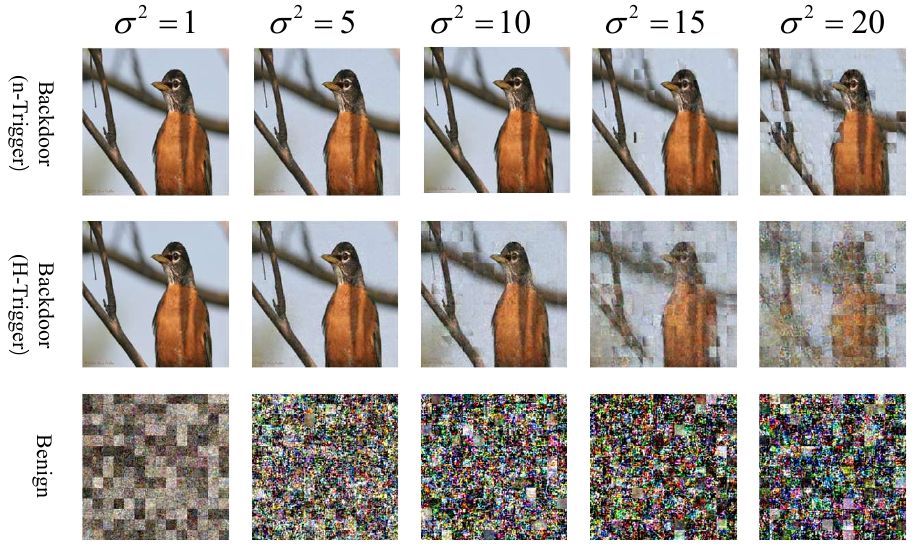}
\caption{ Backdoor detection via noise titration. The backdoored model exhibits pronounced sensitivity to noise, when the variance exceeds a threshold, the decoder reconstructs the predefined target image with high fidelity. In contrast, the benign model fails to produce any structured reconstruction.}
\label{fig_8}
\end{figure*}

\begin{figure*}[t]
\centering
\includegraphics[width=0.95\textwidth]{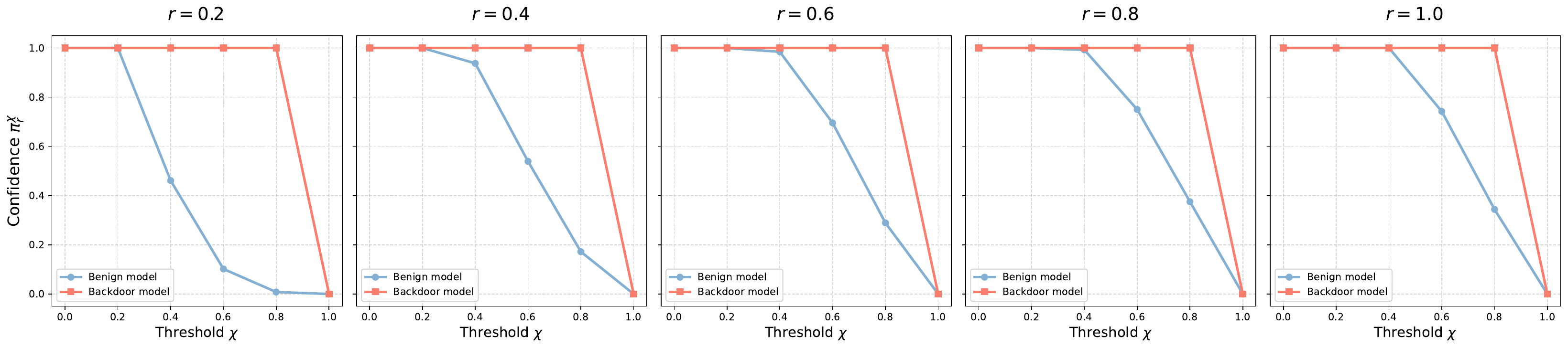}
\caption{ Titration curves vs. threshold $\chi$ for the benign model and backdoor model with different noise intensity $r$.}
\label{fig_9}
\end{figure*}

\subsection{Universal Attack Effectiveness Across SemCom Systems }
We systematically validates the generalizability of CT-BA by evaluating its attack performance across three typical SemCom systems: BDJSCC, ADJSCC, JSCCOFDM, and VIT-MIMO. These systems represent dominant architectures in E2E SemCom systems, covering both block-based and adaptive transmission paradigms. 
\subsubsection{BDJSCC}The first  deep learning (DL) based JSCC system for image source that unifies source compression and channel coding through end-to-end neural network optimization. The architecture employs CNNs for encoder-decoder pair optimization. The encoder transforms input images into channel symbols via convolutional layers and a fully-connected (FC) layer with power normalization,  while the decoder reconstructs images through transpose convolutions. Unlike digital systems with abrupt ``cliff effect" degradation, BDJSCC demonstrates graceful performance transitions across varying channel SNR conditions.
\subsubsection{ADJSCC} An attention DL based JSCC system  that integrates attention feature (AF) module  into CNNs layers. Encoder-decoder pairs use feature learning (FL) modules and AF modules, with FL modules and AF modules are connected alternately. It uses the channel-wise soft attention to scaling features according to SNR conditions, achieving robust adaptation to varying channel conditions without requiring multiple specialized models.
\subsubsection{JSCCOFDM} An E2E JSCC communication system combines trainable CNN layers with nontrainable but differentiable layers representing the multipath channel model and OFDM signal processing blocks. It injecting domain expert knowledge by incorporating OFDM baseband processing blocks into the machine learning framework enhances the overall performance compared to an unstructured CNN.
\subsubsection{VIT-MIMO} A ViT-based JSCC architecture designed for multiple-input multiple-output (MIMO) antenna systems. This system leverages the global self-attention mechanism of transformers to represent semantic features with high discriminability while utilizing SVD to transform the MIMO channel into independent sub-channels for efficient data transmission. By incorporating MIMO physical-layer characteristics into the E2E optimization, this model achieves superior multiplexing and diversity gains compared to single-antenna architectures.
\subsubsection{Results and Discussion}We reproduce BDJSCC, ADJSCC, JSCCOFDM, and VIT-MIMO models using identical hyperparameter configurations from their original implementations. The backdoor models are trained following the procedure outlined in Fig. \ref{fig_4}.
Each model is meticulously designed to address distinct communication challenges: BDJSCC excels in single-SNR environments, ADJSCC is optimized for dynamic SNR variations across a wide range, JSCCOFDM is specifically tailored for robust performance in multipath channels, and VIT-MIMO is optimized for multi-antenna spatial multiplexing. To maintain uniformity, we employ an n-trigger scheme to insert a backdoor into each of the three models, where the bandwidth ratio for all JSCC models is set to R = 1/6 and evaluated on the CIFAR-10 dataset.

As shown in Table \ref{tab:other_system_perfomence}, the $\mathrm{SNR_{train}}$ indicates the training SNR for the model's backdoor task and the main task, the $\mathrm{SNR_{test}}$ indicates the testing SNR for the model's backdoor task and the main task. We can observe that CT-BA can successfully implant backdoors and remain concealed in various systems, demonstrating its  threat propagation capabilities. Notably, even in JSCCOFDM, which employs equalization, and VIT-MIMO, which uses SVD-based sub-channel processing, attackers can bypass these physical-layer signal processing blocks through $n$-trigger attacks. This highlights the diversity and flexibility of the trigger vectors in propagating threats across dominant SemCom architectures.

\subsection{Discussion on Backdoor Detection via Noise Titration}

We first discussed  the applicability of existing state-of-the-art defense mechanisms against the proposed CT-BA. Interpretability-based methods, such as Grad-CAM \cite{selvaraju2017grad}, focus on visualizing specific discriminative regions within the input image to explain model decisions, thereby failing to identify triggers that exist solely as global statistical parameters in the channel layer. Similarly, perturbation-based defenses like STRIP \cite{gao2019strip} operate by superimposing external patterns onto input samples to disrupt trigger coherence, a strategy that proves ineffective against channel-domain triggers dependent on noise power or fading states rather than spatial pixel features. Defense mechanisms like Neural Cleanse \cite{wang2019neural} focus on reverse-engineering static pixel patterns in the input domain, thereby failing to address channel-domain triggers that rely on dynamic statistical parameters rather than fixed spatial features. Consequently, these input-centric strategies fail to capture the attack vectors embedded within the physical communication channel. Activation Clustering \cite{chen2018detecting} detects backdoors by analyzing the neural network activations of training data to separate poisonous samples from legitimate ones based on their distinct cluster distributions. This method relies heavily on the assumption of a classification task where activations can be segmented by discrete class labels to identify multimodal anomalies. In the context of SemCom, which is fundamentally a reconstruction (regression) task with continuous outputs, such label-based segmentation is inapplicable. Fine-Pruning \cite{liu2018fine} mechanism aims to remove backdoors by pruning neurons that are dormant on clean inputs and then fine-tuning the model. However, in our channel-triggered framework, the backdoor neurons activated by specific channel conditions are typically part of the model's legitimate ability to handle channel fading and noise. Pruning these neurons may reduce the system's robustness to natural channel variations and disrupt the main communication tasks.

Inspired by noise response analysis in nonlinear dynamic systems \cite{rosenstein1993practical,poon2001titration}, we propose a noise titration scheme for backdoor detection. Specifically, we inject noise with varying variances into the decoder’s input during inference to probe potential backdoor activation. As illustrated in Fig. \ref{fig_8}, the responses of the benign and backdoored models are compared under controlled noise perturbations. The backdoored model exhibits pronounced sensitivity to noise; when the variance exceeds a threshold, the decoder reconstructs the predefined target image with high fidelity. In contrast, the benign model fails to produce any structured reconstruction.

To quantitatively distinguish these behaviors, we utilize a pre-trained classifier $\Omega$ (e.g., ResNet-18) to compute the high-confidence prediction ratio $\pi_r^{\chi}$. Specifically, we inject i.i.d. normal-distributed noise $\delta \sim N(0,1)$ scaled by intensity $r$ (i.e., $r\delta \sim N(0, r^2)$) into the decoder’s input during inference to probe potential backdoor activation. To characterize the model's response, we record the confidence ratio $\pi_r^{\chi}$ of predictions, defined as:
\begin{equation}
    \pi_r^\chi = \frac{\sum_{i=1}^{N^r} 1_{y > \chi}(\bar{y}_i^r)}{N^r}
\end{equation}
where $\bar{y}_i^r = \max \sigma(\Omega (g_\phi(r\delta_i)))$ is the maximum output of the softmax activation $\sigma(\cdot)$, $1_{y > \chi}(\cdot)$ is the indicator function against a tunable threshold $\chi$, i.e. if $\bar{y}_i^r> \chi$ then $ 1_{y > \chi}(\bar{y}_i^r) =1$, otherwise $ 1_{y > \chi}(\bar{y}_i^r) =0 $, $N^r$ is the number of noise samples. As illustrated in Fig. \ref{fig_9}, we show the changing trend of the confidence of the two kinds of the model as the threshold $\chi$ changes when $r$ takes different values. we see that as $r$ increases, the threshold required for the confidence to decay to  0 becomes larger. In other words, by choosing an appropriate  threshold $\chi$, the confidence of the backdoor model will reach  1 faster than the benign model. We choose $\chi = 0.8$ as the final threshold.



\section{LIMITATIONS AND FUTURE WORKS}

\subsection{SemCom systems use collaborative training}
Distributed training paradigms, particularly Federated Learning, have recently been practically deployed in SemCom systems to enhance data privacy and scalability. However, this collaborative framework introduces significant challenges for backdoor persistence; specifically, the mechanisms of global model aggregation and the iterative nature of multi-round training often lead to the dilution or forgetting of malicious triggers injected by single malicious. This poses a distinct robustness challenge for the proposed CT-BA, as the backdoor task might be forgotten during the average process.

\subsection{Extension to Multi-user and Broadcast Scenarios}
In scenarios with multi-user or broadcast topology, which have a complex structure, the primary challenge lies in the specificity of triggers; in a broadcast channel, a trigger intended for a specific victim may be inadvertently activated by other users experiencing similar channel statistics. A possible solution is to explore higher-dimensional triggers to distinguish between different users. Additionally, managing the trade-off between attack stealthiness and robustness in dynamic networks with multiple interference sources remains a critical hurdle. Potential solutions include adaptive trigger mechanisms that leverage machine learning to track specific channel state transitions, thereby maintaining high attack success rates even in non-stationary multi-user environments.

\subsection{Advanced Defense Mechanisms against Channel Triggers}

Current defense strategies, including the noise titration method discussed in this paper, primarily serve as detection mechanisms rather than proactive mitigation techniques. While noise titration can reveal the presence of a backdoor by probing the decoder's sensitivity, it does not remove the backdoor or recover the benign model. Moreover, sophisticated adversaries may employ "adaptive attacks" by incorporating noise-robustness objectives into the backdoor loss function, potentially rendering simple statistical probing ineffective. Therefore, another pivotal direction for future work is the development of channel-aware unlearning algorithms. We aim to design defense protocols that can not only detect anomalous channel sensitivities but also sanitize the compromised model weights to neutralize the backdoor without degrading the semantic reconstruction quality of the main task.

\section{Conclusion}
To address the significance of understanding the security risks posed to SemCom systems, this paper has introduced CT-BA, a novel backdoor attack paradigm targeting semantic symbols through channel-specific triggers. Unlike conventional input-triggered attacks, CT-BA has leveraged inherent wireless channel parameters as covert triggers, offering enhanced flexibility and feasibility in trigger design. Our evaluation across three benchmark datasets and JSCC systems has demonstrated near-perfect attack success rates while preserving model utility. This work has revealed a new class of backdoor threats in semantic communication, highlighting the need for robust defense mechanisms against channel-aware attacks. Looking ahead, this work necessitates a comprehensive investigation into the security of end-to-end communication systems supporting various modalities (e.g., text, speech, and video), with particular emphasis on exploring potential backdoor attack vectors. The findings will facilitate the development of robust countermeasures to enhance system security.



 
\bibliographystyle{IEEEtran}
\bibliography{IEEEabrv,bare_conf}
%


\section{Biography Section}
 

\begin{IEEEbiography}[{\includegraphics[width=1in,height=1.25in,clip,keepaspectratio]{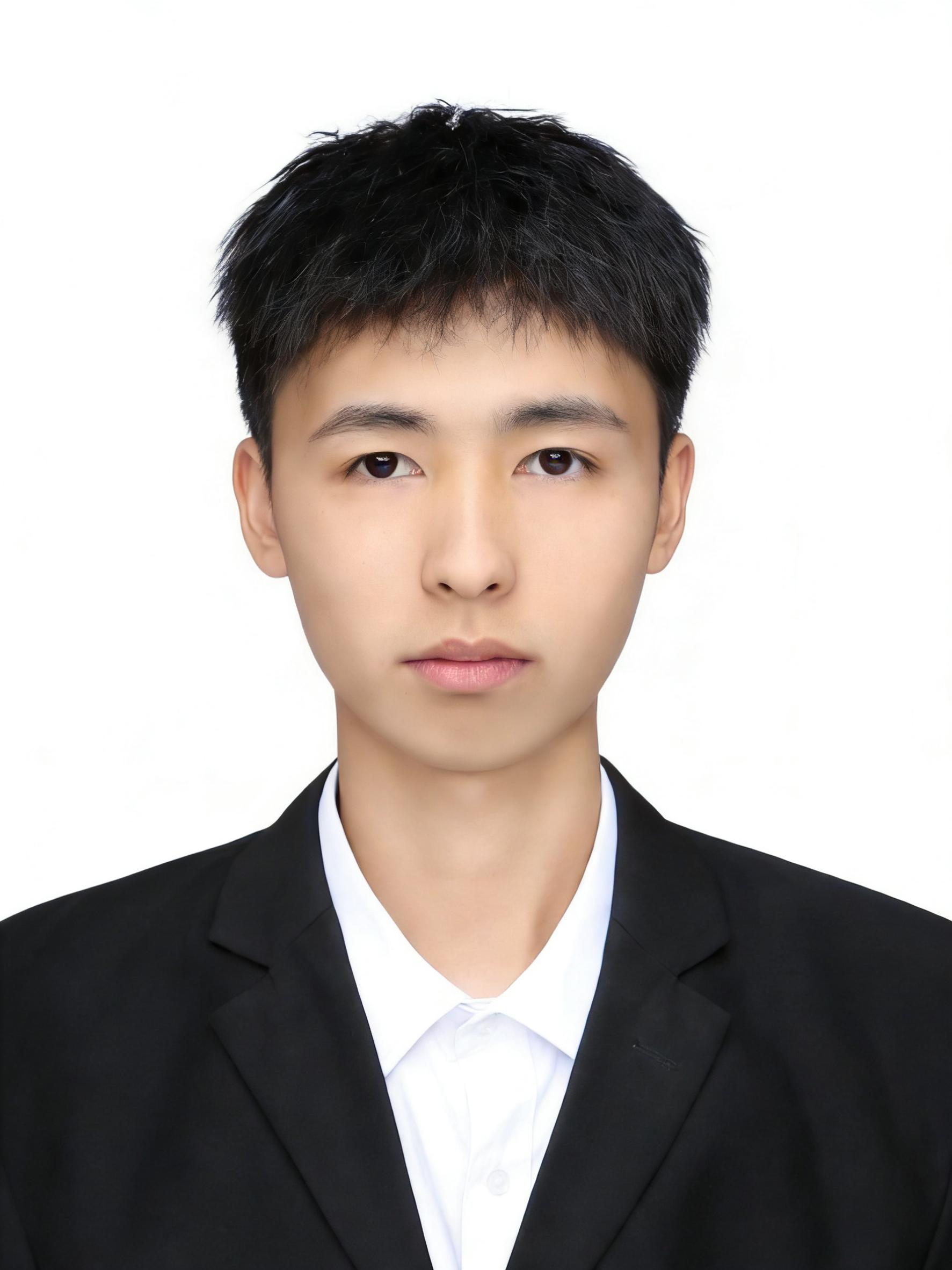}}]{Jialin Wan}
(Student Member, IEEE) is currently pursuing the M.S. degree at Xidian University, Xi’an, China. His current research focuses on intelligent networking and AI security, particularly in the investigation of neural backdoors.
\vspace{-1em}
\end{IEEEbiography}

\begin{IEEEbiography}[{\includegraphics[width=1in,height=1.25in,clip,keepaspectratio]{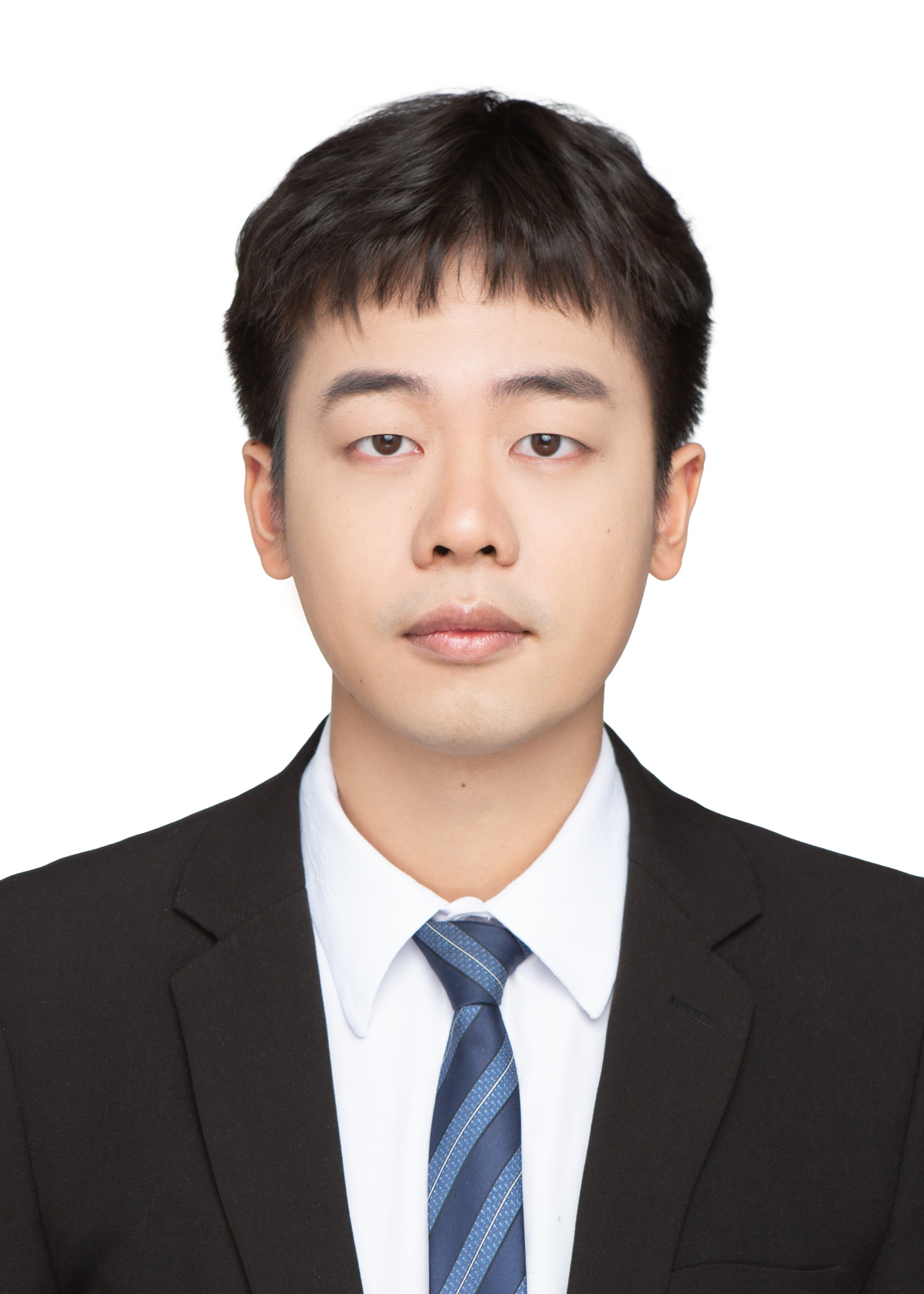}}]{Jinglong Shen}
(Student Member, IEEE) received the B.S. degree in information engineering from Xidian University, Xi’an, China, in 2021. He is currently pursuing the Ph.D. degree in information and communication engineering with the School of Telecommunications Engineering, Xidian University. From September 2025 to September 2026, he was a Visiting Student with the Singapore University of Technology and Design (SUTD), Singapore. His research interests include federated learning, mobile edge computing, and large language models. He was a recipient of the ICCC 2023 Best Demo Award. He has served as a Technical Program Committee (TPC) Member for IEEE GlobeCom 2024-2025 and IEEE ICCC 2024-2025.
\vspace{-1em}
\end{IEEEbiography}

\begin{IEEEbiography}[{\includegraphics[width=1in,height=1.25in,clip,keepaspectratio]{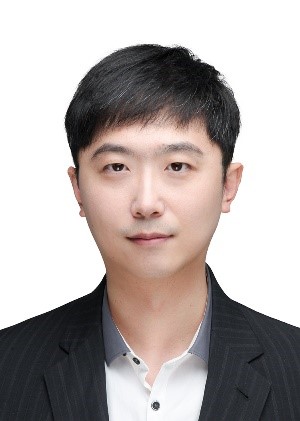}}]{Nan Cheng}
(Senior Member, IEEE) received the Ph.D. degree from the Department of Electrical and Computer Engineering, University of Waterloo in 2016, and B.E. degree and the M.S. degree from the Department of Electronics and Information Engineering, Tongji University, Shanghai, China, in 2009 and 2012, respectively. He worked as a Post-doctoral fellow with the Department of Electrical and Computer Engineering, University of Toronto, from 2017 to 2019. He is currently a professor with the School of Electrical Engineering \& Intelligentization, Dongguan University of Technology and with the State Key Laboratory of ISN and School of Telecommunications Engineering, Xidian University. He has published over 90 journal papers in IEEE Transactions and other top journals. He serves as associate editors for \textit{IEEE Internet of Things Journal}, \textit{IEEE Transactions on Vehicular Technology}, \textit{IEEE Open Journal of Vehicular Technology}, and \textit{Peer-to-Peer Networking and Applications}, and serves/served as guest editors for several journals. His current research focuses on B5G/6G, AI-driven future networks, and space-air-ground integrated network.
\vspace{-1em}
\end{IEEEbiography}

\begin{IEEEbiography}[{\includegraphics[width=1in,height=1.25in,clip,keepaspectratio]{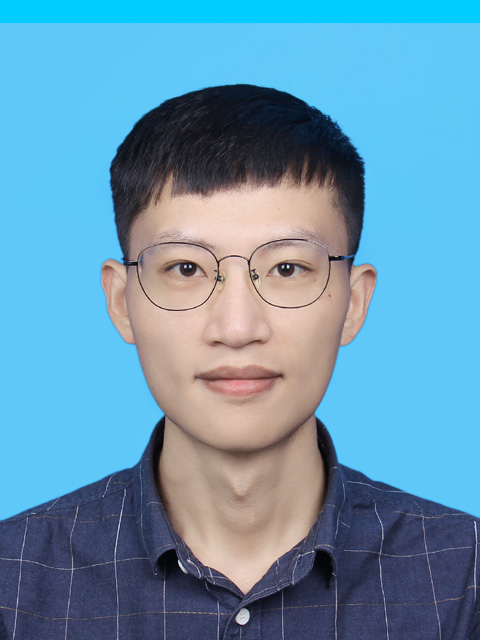}}]{Zhisheng Yin}
(Member, IEEE) received the B.E. degree from the Wuhan Institute of Technology, Wuhan, China, the B.B.A. degree from the Zhongnan University of Economics and Law, Wuhan, in 2012, the M.Sc. degree from the Civil Aviation University of China, Tianjin, China, in 2016, and the Ph.D. degree from the School of Electronics and Information Engineering, Harbin Institute of Technology, Harbin, China, in 2020. From September 2018 to September 2019, he visited with the BBCR Group, Department of Electrical and Computer Engineering, University of Waterloo, Canada. He is currently an Assistant Professor with the School of Cyber Engineering, Xidian University, Xi’an, China. His research interests include space-air-ground integrated networks, wireless communications, cybertwin, and physical layer security.
\vspace{-1em}
\end{IEEEbiography}

\begin{IEEEbiography}[{\includegraphics[width=1in,height=1.25in,clip,keepaspectratio]{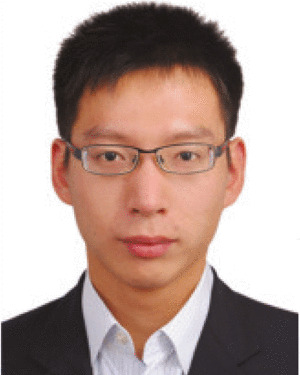}}]{Yiliang Liu}
(Member, IEEE) received the B.E. and M.Sc. degrees in computer science and communication engineering from Jiangsu University, Zhenjiang, China, in 2012 and 2015, respectively, and the Ph.D. degree from the School of Electronics and Information Engineering, Harbin Institute of Technology, Harbin, China, in 2020. He is currently an Associate Professor with the School of Cyber Science and Engineering, Xi’an Jiaotong University, Xi’an, China. His research interests include wireless communication security, physical-layer security, and intelligent connected vehicles. He was the recipient of the Outstanding Doctoral Dissertation Award from China Education Society of Electronics in 2020 and the Best Paper Award from the IEEE Systems Journal in 2021, IEEE VTC-Fall in 2023, IEEE ICITE in 2024, and IEEE AIoT in 2024.
\vspace{-5em}
\end{IEEEbiography}

\begin{IEEEbiography}[{\includegraphics[width=1in,height=1.25in,clip,keepaspectratio]{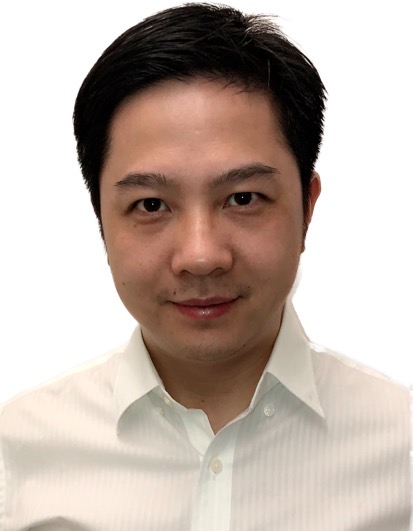}}]{Wenchao Xu}
(Member, IEEE) received the BE and ME degrees in telecommunications engineering from Zhejiang University, Hangzhou, China, in 2008 and 2011, respectively, and the PhD degree in electrical and computing engineering from the University of Waterloo, Canada, in 2018. He is a research assistant professor with The Hong Kong Polytechnic University. In 2011, he joined Alcatel Lucent Shanghai Bell Company Ltd., where he was a software engineer for telecom virtualization. He has also been an assistant professor with the School of Computing and Information Sciences in Caritas Institute of Higher Education, Hong Kong. His research interests includes wireless communication, Internet of Things, distributed computing, and AI enabled networking.
\vspace{-5em}
\end{IEEEbiography}

\begin{IEEEbiography}[{\includegraphics[width=1in,height=1.25in,clip,keepaspectratio]{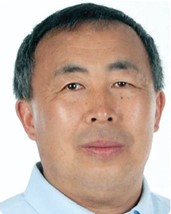}}]{Xuemin (Sherman) Shen}
(Fellow, IEEE) received the PhD degree in electrical engineering from Rutgers University, New Brunswick, NJ, USA, in 1990. He is a University professor with the Department of Electrical and Computer Engineering, University of Waterloo, Canada. His research focuses on network resource management, wireless network security, Internet of Things, AI for networks, and vehicular networks. He is a registered professional engineer of Ontario, Canada, an Engineering Institute of Canada fellow, a Canadian Academy of Engineering fellow, a Royal Society of Canada fellow, a Chinese Academy of Engineering foreign member, an International fellow of the Engineering Academy of Japan, and a distinguished lecturer of the IEEE Vehicular Technology Society and Communications Society.

Dr. Shen received “West Lake Friendship Award” from Zhejiang Province in 2023, President’s Excellence in Research from the University of Waterloo in 2022, the Canadian Award for Telecommunications Research from the Canadian Society of Information Theory (CSIT) in 2021, the R.A. Fessenden Award in 2019 from IEEE, Canada, Award of Merit from the Federation of Chinese Canadian Professionals (Ontario) in 2019, James Evans Avant Garde Award in 2018 from the IEEE Vehicular Technology Society, Joseph LoCicero Award in 2015 and Education Award in 2017 from the IEEE Communications Society (ComSoc), and Technical Recognition Award from Wireless Communications Technical Committee (2019) and AHSN Technical Committee (2013). He is the past president of the IEEE Communications Society. He was the vice president for Technical \& Educational Activities, vice president for Publications, member-at-large on the Board of Governors, chair of the Distinguished Lecturer Selection Committee, and member of IEEE Fellow Selection Committee of the ComSoc. He served as the editor-in-chief of IEEE Internet of Things Journal, IEEE Network, and IET Communications.
\end{IEEEbiography}


\end{document}